\documentclass[traditabstract, hidelinks]{aa} 
\usepackage{graphicx}

\usepackage{lscape}
\usepackage{amssymb}
\usepackage{amsmath}

\usepackage{txfonts}
\usepackage{multirow}
\usepackage{booktabs}
\usepackage{subfigure}
\usepackage{placeins}
\usepackage{verbatim}
\usepackage{adjustbox}
\usepackage[toc,page]{appendix} 
\usepackage{tabularx}
\usepackage{lscape}
\usepackage{textcomp}
\usepackage{gensymb}
\usepackage{hyperref}
\usepackage{multicol}
\usepackage[symbol]{footmisc}
\usepackage{tikz}
\usepackage{caption}
\usepackage{longtable}
\usepackage{xtab}

\usepackage{natbib}

\usepackage{color}

\newcommand{\bd}{\begin{displaymath}}
\newcommand{\ed}{\end{displaymath}}
\newcommand{\be}{\begin{equation}}
\newcommand{\ee}{\end{equation}}
\newcommand{\beaa}{\begin{eqnarray*}}
\newcommand{\eeaa}{\end{eqnarray*}}
\newcommand{\bea}{\begin{eqnarray}}
\newcommand{\eea}{\end{eqnarray}}

\def\HST{{HST}{}}
\def\MUSE{{MUSE}{}}
\def\VLT{{VLT}{}}
\def\JWST{{JWST}{}}
\def\GLEE{\textsc{Glee}}

\def\rms{0.39\arcsec}          
\def\musecat{162}              

\defcitealias{grillo16}{G16}
\begin{document}

   \title{Improved model of the Supernova Refsdal cluster MACS~J1149.5+2223 thanks to \textit{VLT}/MUSE\thanks{This work is based in large part on data collected at ESO VLT (prog. IDs 294.A-5032 and 105.20P5.001) and NASA HST.}}
  \titlerunning{Improved model of MACS~J1149 }

   \author{S.~Schuldt\inst{1}\inst{,2} \and
            C.~Grillo\inst{1}\inst{,2} \and 
            G.~B.~Caminha\inst{3}\inst{,4} \and
            A.~Mercurio\inst{5}\inst{,6}\inst{,7} \and
            P.~Rosati\inst{8}\inst{,9}\and
            T.~Morishita\inst{10}\and
            M.~Stiavelli\inst{11}\and
            S.~H.~Suyu\inst{3}\inst{,4}\inst{,12}\and
            P.~Bergamini\inst{1}\inst{,9}\and
            M.~Brescia\inst{13}\and
            F.~Calura\inst{9}\and
            M.~Meneghetti\inst{9}
                        }

        \institute{Dipartimento di Fisica, Universit\`a  degli Studi di Milano, via Celoria 16, I-20133 Milano, Italy\\
        e-mail: \href{mailto:stefan.schuldt@unimi.it}{\tt stefan.schuldt@unimi.it}
        \and
        INAF -- IASF Milano, via A. Corti 12, I-20133 Milano, Italy
        \and
        Technical University of Munich, TUM School of Natural Sciences, Department of Physics,  James-Franck-Stra{\ss}e 1, 85748 Garching, Germany
        \and
        Max-Planck-Institut f{\"u}r Astrophysik, Karl-Schwarzschild Stra{\ss}e 1, 85748 Garching, Germany
        \and
        Università di Salerno, Dipartimento di Fisica ``E.R. Caianiello'', Via Giovanni Paolo II 132, I-84084 Fisciano (SA), Italy
        \and
        INAF -- Osservatorio Astronomico di Capodimonte, Via Moiariello 16, I-80131 Napoli, Italy
        \and
        INFN – Gruppo Collegato di Salerno - Sezione di Napoli,  Dipartimento di Fisica "E.R. Caianiello", Università di Salerno, via Giovanni Paolo II, 132 - I-84084 Fisciano (SA), Italy
        \and
        Dipartimento di Fisica e Scienze della Terra, Universit\`a degli Studi di Ferrara, via Saragat 1, I-44122 Ferrara, Italy
        \and
        INAF -- OAS, Osservatorio di Astrofisica e Scienza dello Spazio di Bologna, via Gobetti 93/3, I-40129 Bologna, Italy 
        \and
        IPAC, California Institute of Technology, MC 314-6, 1200 E. California Boulevard, Pasadena, CA 91125, USA
        \and
        Space Telescope Science Institute, 3700 San Martin Drive, Baltimore, MD 21218, USA
        \and
        Academia Sinica Institute of Astronomy and Astrophysics (ASIAA), 11F of ASMAB, No.1, Section 4, Roosevelt Road, Taipei 10617, Taiwan
        \and
        Dipartimento di Fisica ``E. Pancini", Università degli Studi di Napoli Federico II, Via Cinthia, 21, I-80126 Napoli, Italy
         }

   \date{Received 7 February 2024 / Accepted 17. May 2024}

 
  \abstract{
  We present new \VLT/\MUSE \, observations of the \textit{Hubble} Frontier Field (HFF) galaxy cluster MACS J1149.5+2223, lensing the well-known supernova ``Refsdal'' into multiple images, which has enabled the first cosmological applications with a strongly lensed supernova. Thanks to these data, targeting a northern region of the cluster and thus complementing our previous \MUSE \, program on the cluster core, we have released a new catalog containing \musecat \, secure spectroscopic redshifts. We confirmed 22 cluster members, which had previously been only photometrically selected, and detected ten additional ones, resulting in a total of 308 secure members, of which 63\% are spectroscopically confirmed. We further identified 17 new spectroscopic multiple images belonging to six different background sources. By exploiting these new and our previously published \MUSE \, data, in combination with the deep HFF images, we developed an improved total mass model of MACS J1149.5+2223. This model includes 308 total mass components for the member galaxies and requires four additional mass profiles, one of which is associated with a cluster galaxy overdensity identified in the north, representing the dark matter mass distribution on larger scales. The values of the resulting 34 free parameters are optimized based on the observed positions of 106 multiple images from 34 different families, that cover an extended redshift range between 1.240 and 5.983. Our final model has a multiple image position root mean square value of \rms, which is in good agreement with other cluster lens models based on a similar number of multiple images. With this refined mass model, we have paved the way toward an improved strong-lensing analyses that will exploit the deep and high resolution observations with \HST \, and \JWST \, on a pixel level in the region of the supernova Refsdal host. This will increase the number of observables by around two orders of magnitude, thus offering the opportunity to carry out more precise and accurate cosmographic measurements in the future.
  }

   \keywords{gravitational lensing: strong $-$ methods: data analysis $-$ catalogs $-$ galaxies: clusters: general $-$ galaxies:distances and redshifts $-$ galaxies: clusters: individuals: MACS J1149.5$+$2223}

   \maketitle

\section{Introduction}
\label{sec:intro}

Gravitational lensing describes the effect of a massive object, such as a galaxy cluster, deflecting the light coming from background sources. It enables us to probe several different aspects of the Universe. Since the deflection depends on the total mass of a lens, that is, both the dark matter and baryonic components, gravitational lensing can be used to study its dark matter mass distribution \cite[e.g.,][]{grillo15, limousin16, cerny18, schuldt19, Meneghetti20, Meneghetti22, Meneghetti23, shajib21, wang22}. Furthermore, it allows the study of the evolution of galaxy clusters \citep[e.g.,][]{annunziatella16, mercurio21} and faint high-redshift galaxies that are observable thanks to the lensing magnification effect \citep[e.g.,][]{coe13, calcura21, vanzella21, mestric22}.

\begin{figure*}[t!]
    \centering
    \includegraphics[trim={103 38 80 40},clip, width=\linewidth]{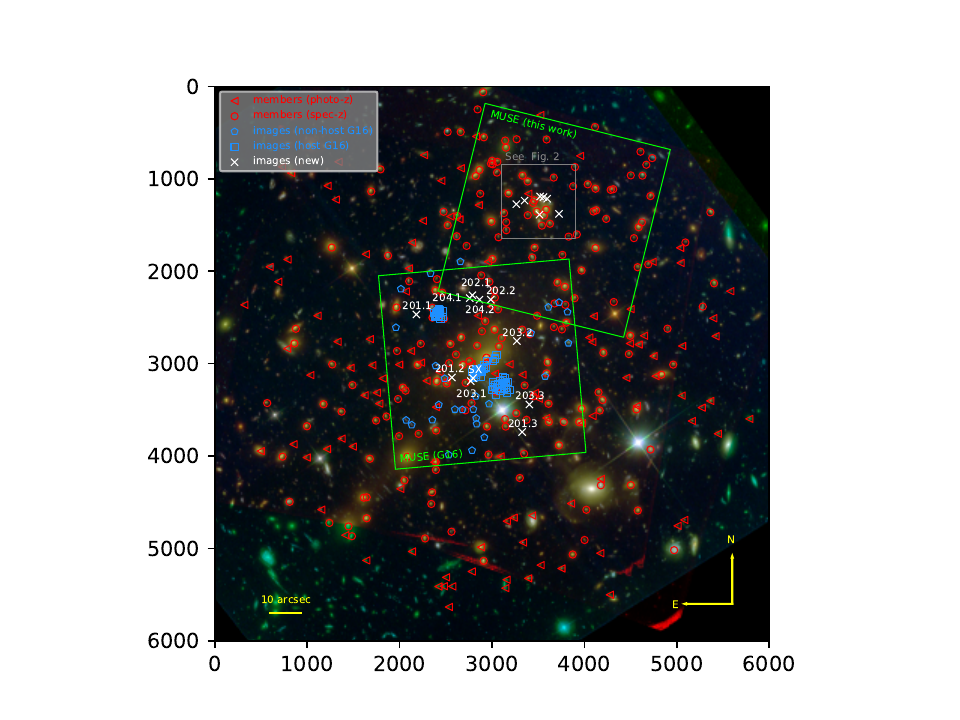}
    \caption{Color-composite image of MACS~1149 obtained by combining the HFF observations  (F435W + F465W for blue, F555W + F606W + F625W + F775W + F814W + F850LP for green, and F105W + F110W + F125W + F140W + F160W for red). The cluster members (red) either photometrically selected (triangle) or spectroscopically confirmed (circle), and the multiple image systems are shown. We distinguish the multiple images between those previously known from \citetalias{grillo16} (blue), either belonging to the SN Refsdal host (square) or at other redshifts (pentagon), and those newly identified  in this work (white). We highlight the regions observed with \MUSE \, (green) and the zoom-in region (gray) shown in Fig.~\ref{fig:HFF_zoom}}
    \label{fig:HFF}
\end{figure*}

In addition, in the rare case of a multiply imaged time-variable source, such as a quasar or a supernova (SN), lensing gives us a unique opportunity to exploit these lensing systems for cosmological applications, including the measurement of the Hubble constant $H_0$ and of the geometry of the Universe \citep[see][and references therein]{treu22, suyu24}. This technique is called time-delay cosmography. 
To date, SN ``Refsdal'' \citep[$z = 1.489$,][]{kelly15_ATel, kelly16a} is the only strongly lensed supernova used for precise time-delay cosmography \citep[e.g.,][]{grillo18, grillo20, grillo24, kelly23a}, making the lens cluster a unique and well studied field. 
Given the increasing number of detected SNe strongly lensed by galaxy clusters \citep[e.g.,][]{kelly15, rodney16, rodney21, frye24, pierel24}, where the time delays are on the order of months to years, compared to those of days to weeks of galaxy-scale systems, detailed total mass models of these lens clusters will become crucial for time-delay cosmography.

SN Refsdal is multiply imaged by the \textit{Hubble} Frontier Field (HFF) galaxy cluster MACS~J1149.5+2223 \citep[hereafter, MACS~1149;][hereafter \citetalias{grillo16}]{treu16, grillo16} located at a redshift of $z_\text{c}=0.5422$. Given that it is in the HFF cluster sample, as well as being part of the Cluster Lensing And Supernova survey with \textit{Hubble} \citep[CLASH,][]{postman12} and the Grism Lens-Amplified Survey from Space \citep[GLASS,][]{treu15, schmidt14}, this cluster is equipped with extensive spectroscopic and photometric data. This set of data is complemented by deep Multi Unit Spectroscopic Explorer \citep[\MUSE,][]{bacon12} observations of the cluster core carried out by \citetalias{grillo16} to securely select cluster members and multiple image systems for a robust strong lensing model (see Fig.~\ref{fig:HFF}).

\begin{figure}[t!]
    \centering
    \includegraphics[trim={103 38 90 40},clip, width=\linewidth]{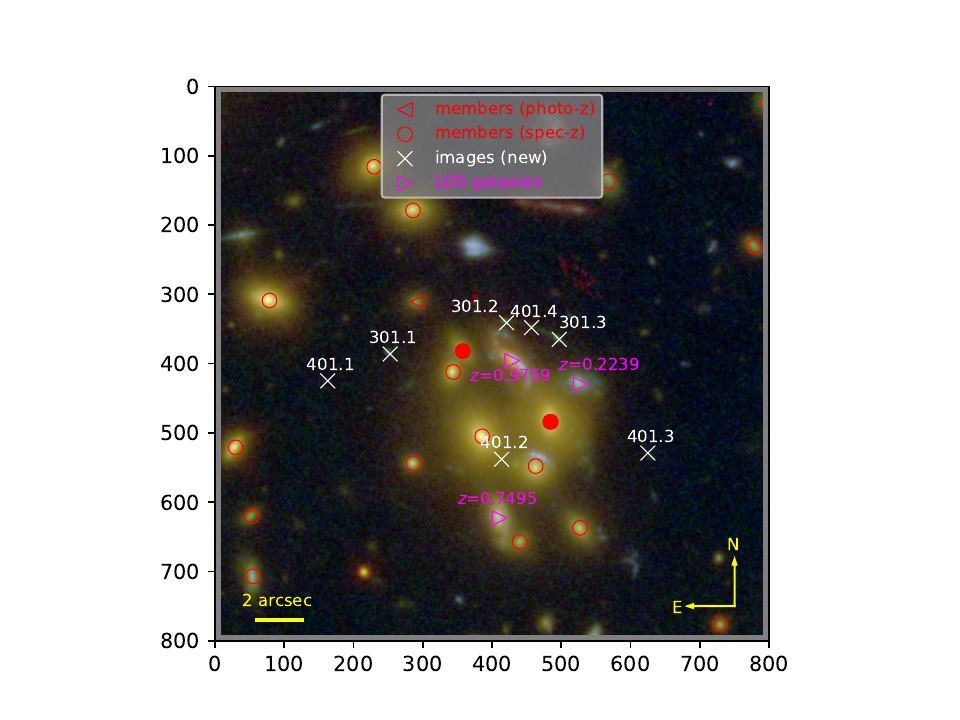}
    \caption{Color-composite image (as in Fig.~\ref{fig:HFF}) of the galaxy group in the northern field of MACS~1149. Shown are the cluster members (red) and the multiple image systems (white). Furthermore,  we indicate three line-of-sight galaxies (magenta), with the corresponding redshift values, and two cluster members (red-filled circles) whose effect on the total lens mass model was tested (see Sect.~\ref{sec:modeling:tests} for more details).}
    \label{fig:HFF_zoom}
\end{figure}

To further expand the data set of this peculiar lens cluster and to improve on the strong lensing model, we present additional 5.5 hours of \MUSE \, observations (PI A.~Mercurio) from the Very Large Telescope (\VLT) of the European Southern Observatory. These observations were carried out in 2022 and 2023, and targeted a region in the northern part of the cluster (see Fig.~\ref{fig:HFF_zoom}). Therefore, these observations perfectly complement our previous \MUSE data, enabling us to significantly extend our redshift catalog with new spectroscopic redshifts. They also reveal several new multiply lensed sources and spectroscopically confirms numerous cluster members, that had previously only been photometrically identified.

By exploiting these new \MUSE \, observations, complemented with existing high-quality spectroscopic and photometric data, we present an improved total mass model of the cluster MACS~1149. We incorporated all newly identified multiple images and take all 308 securely identified cluster members into account. We herewith publicly release the presented cluster lens model, which is poised to play a fundamental role in improving time-delay cosmography and in studying the intrinsic properties of high-redshift sources \citep[see, e.g.,][]{stiavelli23, morishita24a, morishita24b}.

The paper is organized as follows. We introduce the \HST \, data in Sect.~\ref{sec:hst}, and the spectroscopic data obtained with \MUSE in Sect.~\ref{sec:muse}, along with the latest redshift catalog of MACS~1149 and a discussion of the newly discovered multiply imaged sources. In Sect.~\ref{sec:modeling}, we describe our strong lensing mass model of the cluster and the tested mass parametrizations. The results and conclusions are given in Sect.~\ref{sec:conclusion}.

Throughout this paper, we assume a flat $\Lambda$CDM cosmology with $H_0=70 \, \text{km s}^{-1}\,\text{Mpc}^{-1}$ and $\Omega_\text{M}=1-\Omega_\text{DE}=0.3$. In this cosmology, 1\arcsec\, corresponds to 6.36~kpc at the cluster redshift \mbox{$z_\text{c}=0.5422$}. All magnitudes are given in the AB system \citep{oke83,fukugita96}.

\FloatBarrier
\section{\HST \, imaging}
\label{sec:hst}

The cluster MACS~1149 was observed as part of the CLASH (PI M.~Postman) and HFF (PI J.~Lotz) programs with more than 50 \HST \, orbits in 15 broadband filters. These data, complemented with previously obtained images \citep[see][]{postman12, zheng12, jouvel14}, were processed with standard calibration techniques and drizzled to mosaics \citep[see e.g.,][]{koekemoer07, koekemoer11}. In this work, we use all available filters of the HFF stacks with a pixel size of 0.03\arcsec, shown in Fig.~\ref{fig:HFF}, to confirm multiple image systems (see Sect.~\ref{sec:muse:analysis:images}) and the HFF F160W image for the cluster member selection (see Sect.~\ref{sec:muse:analysis:members}).

\FloatBarrier
\section{\VLT/\MUSE \ spectroscopy}
\label{sec:muse}

MACS~1149 was observed with the \VLT/\MUSE \, instrument in two campaigns. The first data set was obtained in 2015 under the ESO programme ID 294.A-5032 (PI C.~Grillo) and covered the core of the cluster, including the SN Refsdal host galaxy, for a total exposure time of 4.8 hours. As stated in \citetalias{grillo16}, the observational conditions were clear and photometric with a seeing of 0.81\arcsec.

Additional 5.5 hours of \MUSE \, observations, centered on the north-west region of the cluster (see Fig.~\ref{fig:HFF}), were obtained in 2022 and 2023 under the programme ID 105.20P5 (PI A.~Mercurio). The data were taken in clear sky and good seeing (0.76\arcsec) conditions. In the following, we present the data reduction procedure and resulting redshift catalog.

\subsection{Data reduction}
\label{sec:muse:reduction}

As described in \citetalias{grillo16}, the \MUSE \, observations from 2015 were reduced with the \MUSE \, data reduction software version 1.0 in 2015. Given the small overlap between this \MUSE \, pointing and the new observations taken in 2022 and 2023 (see Fig.~\ref{fig:HFF}), we performed a new, combined data reduction with the pipeline version 2.8.5. We followed the standard procedure, such as bias subtraction, flat-fielding, wavelength calibration, and exposure map alignment. We refer to \citetalias{grillo16} and \citet{2016A&A...585A..27K} for further details. 

The \MUSE \, datacube was aligned with the \HST \, images, using as references the HST F606W band and the \MUSE \, ``white image". We degraded the \HST \, image to the \MUSE \, white image resolution and employed the positions of 94 compact sources to match both data sets. The root mean square (RMS) separation of the final matched data sets is $0.054\arcsec$, thus significantly below the \MUSE \, pixel size of $0.2\arcsec$. Furthermore, we found no evidence of relative rotation between \MUSE \, and \HST.

\subsection{Spectral analysis}
\label{sec:muse:analysis}

Since the \MUSE \, observations from 2015 were already analyzed, only each extracted spectrum from the new \MUSE \, data, including a re-inspection of objects in the area covered by both \MUSE \, pointings, was inspected individually. As already done in previous \MUSE \, redshift catalogs, we assign to each redshift value a quality flag (QF) that indicates its reliability.  A QF of 3 corresponds to a very secure ($\delta z < 0.001$) measurement, while a QF of 2 corresponds to a secure ($\delta z < 0.01$) one, and of 1 to an unsecure ($\delta z \geq 0.01$) one. A QF of 9 is provided for redshifts that are based on a single emission line, and, since the \MUSE \, spectral resolution allows us to distinguish the shape or doublet nature of narrow emission lines (e.g., Lyman-$\alpha$ and O$_\text{II}$), considered as secure. These error estimates include systematic uncertainties due to different methods of redshift measurements. Following these criteria, we have \musecat \, secure (i.e., QF $>1$) redshifts, which we list in Table~\ref{tab:z_muse}. Given the redshift uncertainty, the spectral resolution of \MUSE, and the number of identified lines in the spectra, we follow our previous publications and provide redshift values with four digits after the decimal point for foreground objects and cluster members, but with three digits for background sources.

\begin{figure}[t!]
    \centering
    \includegraphics[width=\linewidth]{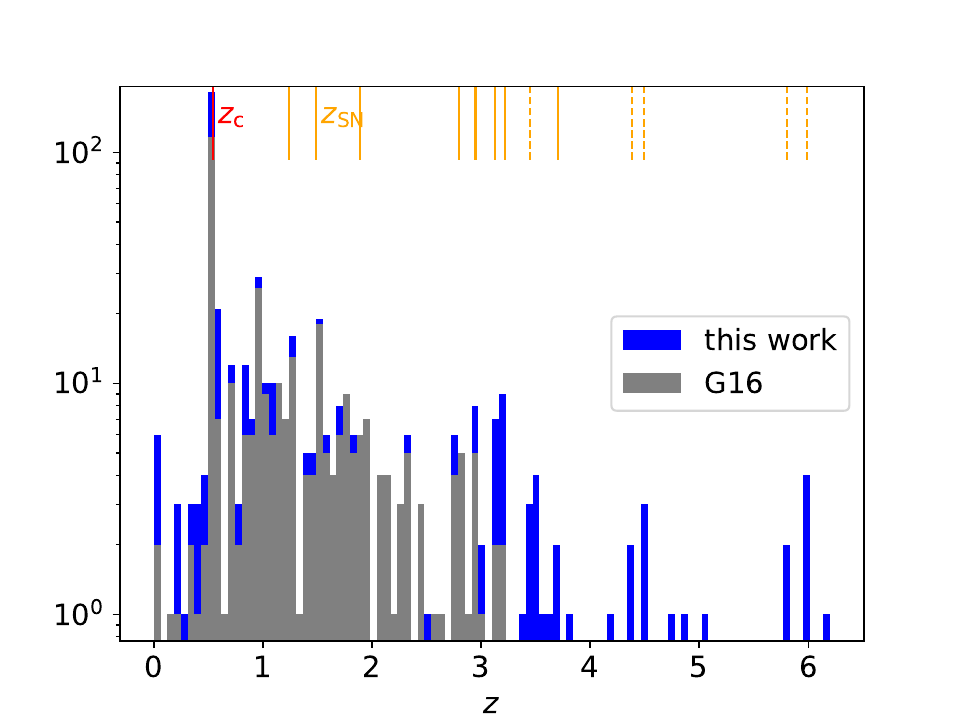}
    \caption{Stacked histogram of the redshift distribution in the MACS~1149 field, divided into previously known (gray, \citetalias{grillo16}) and new or updated\protect\footnote \, (blue) redshifts. As a reference, the redshift of the cluster $z_\text{c}=0.5422$ is indicated by a red line and that of multiply lensed sources with orange lines (solid and dashed lines for the previously known and new identifications, respectively), including SN Refsdal at $z_\text{SN}=1.489$.
    }
    \label{fig:redshifts}
\end{figure}

The secure redshifts listed in Table~\ref{tab:z_muse} are integrated in the previous redshift catalog from \citetalias{grillo16}, which contains now 509 spectroscopic redshifts 
and is made available in electronic form with this publication\footnote{We release the redshift catalog at the following link upon publication: \url{https://www.fe.infn.it/astro/lensing/}}. We show a histogram of the full redshift catalog in Fig.~\ref{fig:redshifts}, divided into redshifts previously published by \citetalias{grillo16} (gray) and new redshifts (blue). This highlights the significant increase in the covered redshift range, as we have now several secure redshifts up to $z\sim6$. 

\FloatBarrier
\footnotetext{For example, the redshift of the multiple images 13261 and 13457 at $z=3.704$ (listed as family 14 in Table~2 of \citetalias{grillo16}).}

\subsubsection{Cluster members}
\label{sec:muse:analysis:members}

To select the cluster members of MACS~1149, we adopted the same selection criteria as in \citetalias{grillo16}. In short, we selected all galaxies that have a secure redshift in the range $ 0.520 \leq z \leq 0.570$, which are either spectroscopically confirmed with \MUSE \, \citepalias[and this work]{grillo16} or obtained from GLASS \citep{treu16}. These spectroscopically confirmed cluster members are, where possible, centered on their luminosity centroids on the HFF images; otherwise the coordinates of the spectroscopic detection itself are used. This spectroscopic sample is complemented with robust photometrically selected members as in \citetalias{grillo16}.

\begin{figure}[t!]
    \centering
    \includegraphics[trim={30 8 30 40},clip, width=\linewidth]{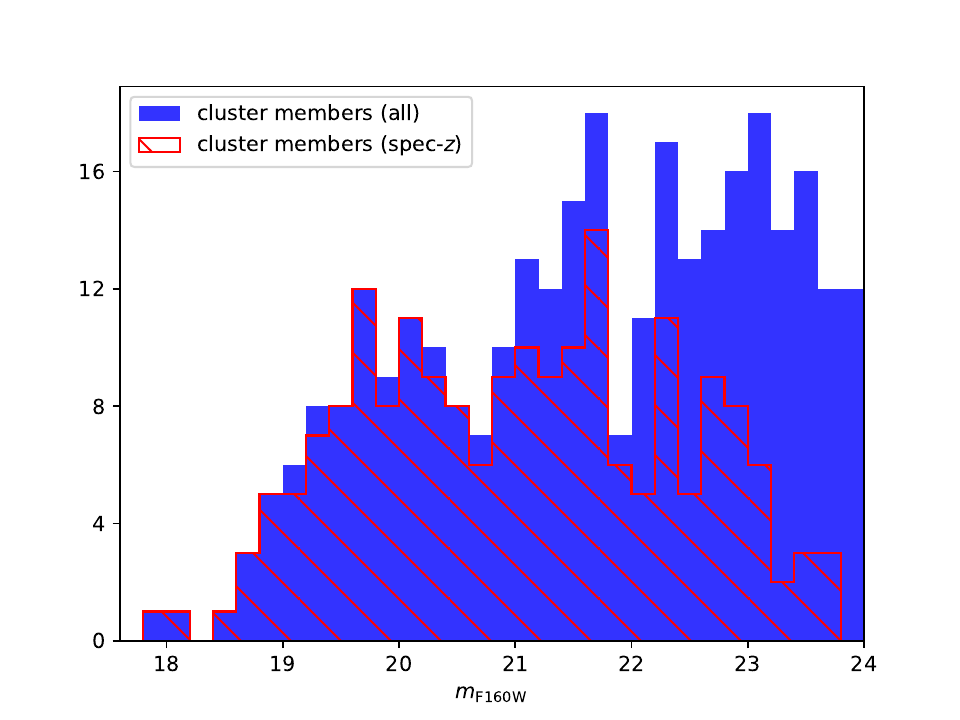}
    \caption{Histogram of the cluster member magnitudes in the HFF F160W band.}
    \label{fig:hist_mag}
\end{figure}

Since the total mass of a cluster member is assumed to scale according to the galaxy total luminosity in the F160W band (see Sect.~\ref{sec:modeling:cm}), we included in the cluster total mass model only the most massive cluster members, namely, those that are brighter than $m_\text{F160W}=24$, as done in \citetalias{grillo16}. This frequently adopted limit is well below the threshold of $m_\text{F160W} \leq 21$, identified by \citet{bergamini23b}, after which no significant difference in the accuracy of a cluster total mass reconstruction was observed. This is also in agreement with \citet{raney21}, who have tested the same threshold in the F814W band. To measure the galaxy F160W total magnitudes, we adopted Kron magnitudes, which approximate those extracted from GALFIT \citep{peng02} very well \citepalias{grillo16}, resulting in reliable and unbiased total magnitudes. For this purpose, we favored the image from the HFF program over the shallower CLASH image used in \citetalias{grillo16}. As a consequence, we obtained slightly different magnitudes and excluded now one cluster member (ID 143 in Table~6 of \citetalias{grillo16}) as it was fainter than $m_\text{F160W}=24$. In contrast, we included six new members, in the outer regions of the cluster, which fall now inside the \HST \, FoV. The distribution of the cluster member magnitudes is shown in Fig.~\ref{fig:hist_mag}.

Thanks to the new \MUSE \, observations, we were able to spectroscopically confirm 22 cluster members that were previously only photometrically selected. Furthermore, we rejected one cluster member, making it to a  photometric outlier (member ID 46 in Table~6 of \citetalias{grillo16}), while four objects (member IDs 305, 306, 308, 309) that were previously excluded from the member sample, based on the spectroscopic redshifts from GLASS, are now included among the cluster members. These four members are relatively faint ($m_\text{F160W} > 22.6$), possibly explaining the wrong measurement in \citet{treu16}. Given their low total luminosity, we expect that excluding them in previously presented cluster mass models would affect them only slightly.

In summary, we were able to securely identify 308 cluster members, containing in total 195 spectroscopically confirmed members (i.e., 63.3\%); of these, ten were entirely new members. The members are listed in Table~\ref{tab:clustermembers} and shown in Fig.~\ref{fig:HFF} (in red). The parametrization of the cluster members in the total mass model and the impact of the changes compared to \citetalias{grillo16} are presented in Sect.~\ref{sec:modeling}.

\begin{table}[t!]
    \caption{Cluster members of MACS~1149 considered in our mass model.}
    \begin{center}
    \begin{tabular}{ccccc}
    \hline \hline \noalign{\smallskip}
ID& RA & Dec. & F160W &  Sel. criterion \\  & (J2000) & (J2000) & [mag] & \\ \hline
1 & 177.37311 & 22.39300 & $21.17 \pm 0.01$ & P \\ 
2 & 177.37497 & 22.40452 & $23.26 \pm 0.03$ & P \\ 
3 & 177.37618 & 22.39166 & $22.24 \pm 0.01$ & P \\ 
4 & 177.37640 & 22.40045 & $22.26 \pm 0.02$ & P \\ 
5 & 177.37655 & 22.40522 & $20.41 \pm 0.05$ & G \\ 
6 & 177.37663 & 22.40207 & $23.30 \pm 0.04$ & P \\ 
7 & 177.37692 & 22.39462 & $21.20 \pm 0.01$ & P \\ 
8 & 177.37766 & 22.40312 & $21.18 \pm 0.07$ & G \\ 
9 & 177.37776 & 22.39401 & $23.75 \pm 0.03$ & P \\ 
10& 177.37871 & 22.39281 & $22.77 \pm 0.02$ & P \\ 
$\vdots$ & \vdots & \vdots & \vdots & \vdots\\
\noalign{\smallskip}\hline
    \end{tabular}
    \end{center}
     \textbf{Notes.} Columns give, from left to right, the IDs following Table~6 of \citetalias{grillo16}, right ascension, declination, and the magnitude from the HFF F160W image with the corresponding uncertainty. In the last column, we indicate whether it was selected photometrically (P), based on the WFC3-IR-GRISM redshift (G) from GLASS \citep{treu16}, the \MUSE \, observations presented in \citetalias{grillo16} (M1), or in this publication (M2). This catalog is downloadable in its entirety from the web\footnote{This catalog is downloadable in its entirety from \url{https://www.fe.infn.it/astro/lensing/}}.
    \label{tab:clustermembers}
\end{table}

\subsubsection{Multiply lensed sources}
\label{sec:muse:analysis:images}

\begin{figure*}[t!]
  \includegraphics[width = 0.66\columnwidth]{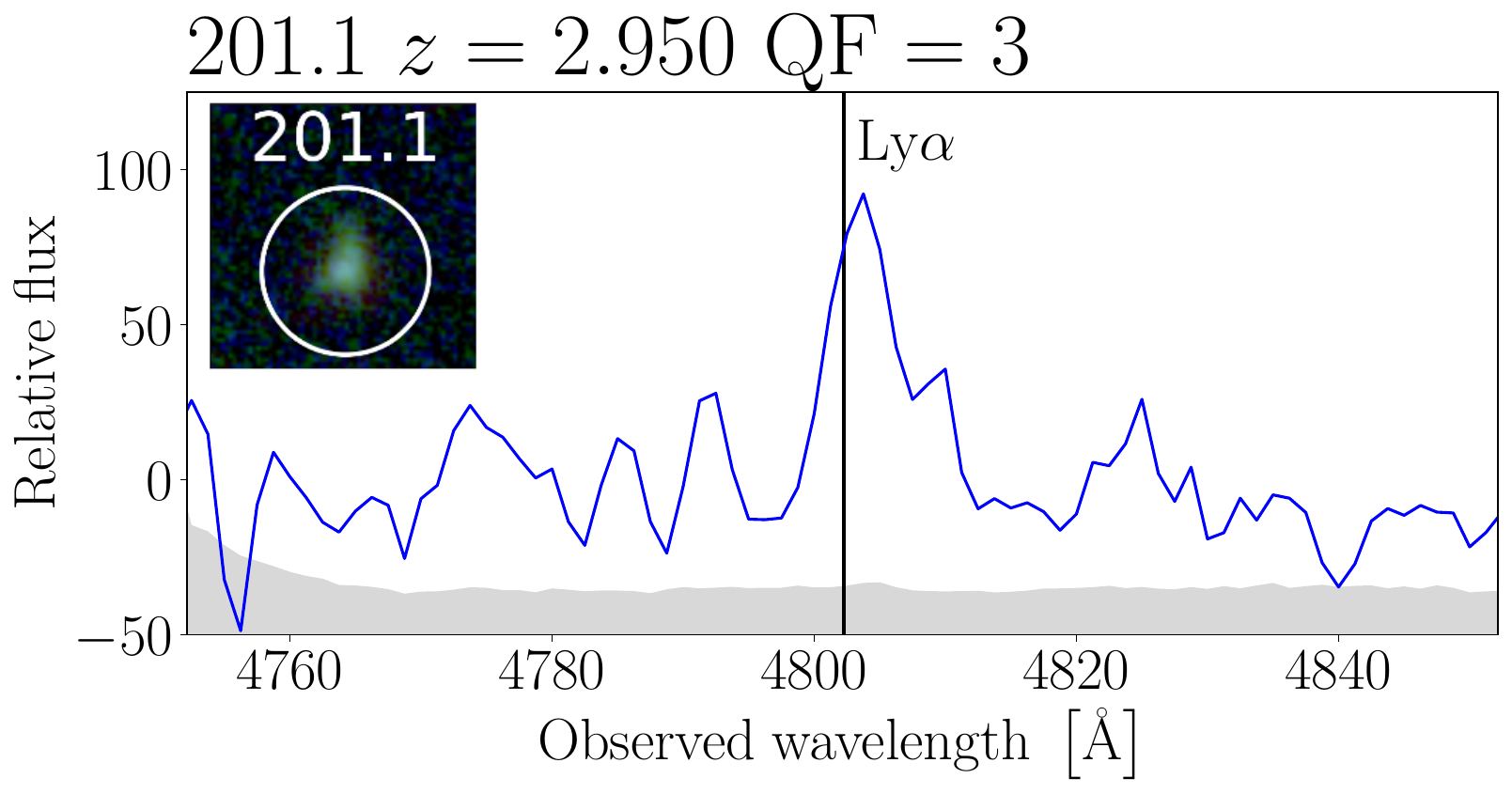}
  \includegraphics[width = 0.66\columnwidth]{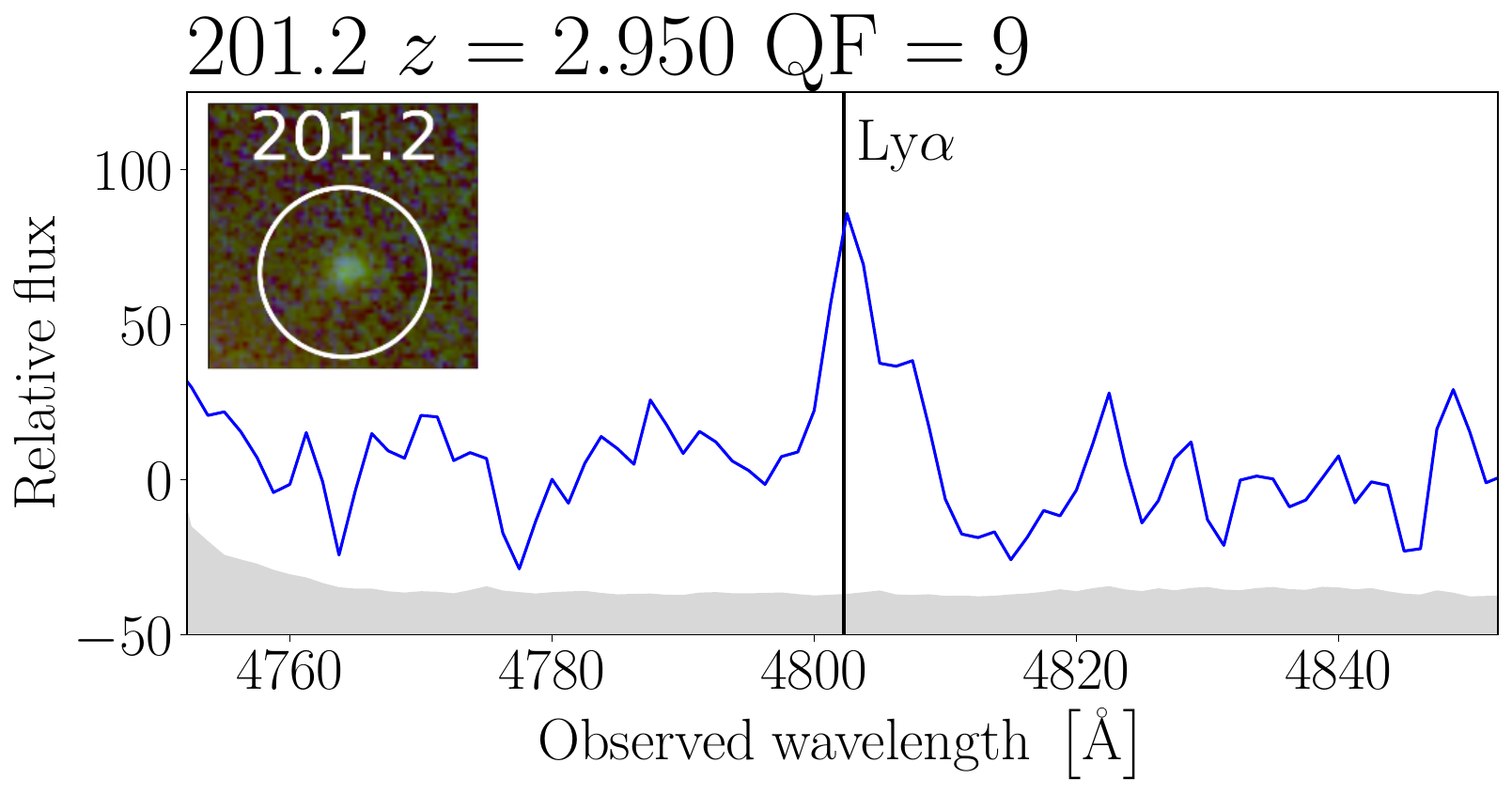}
  \includegraphics[width = 0.66\columnwidth]{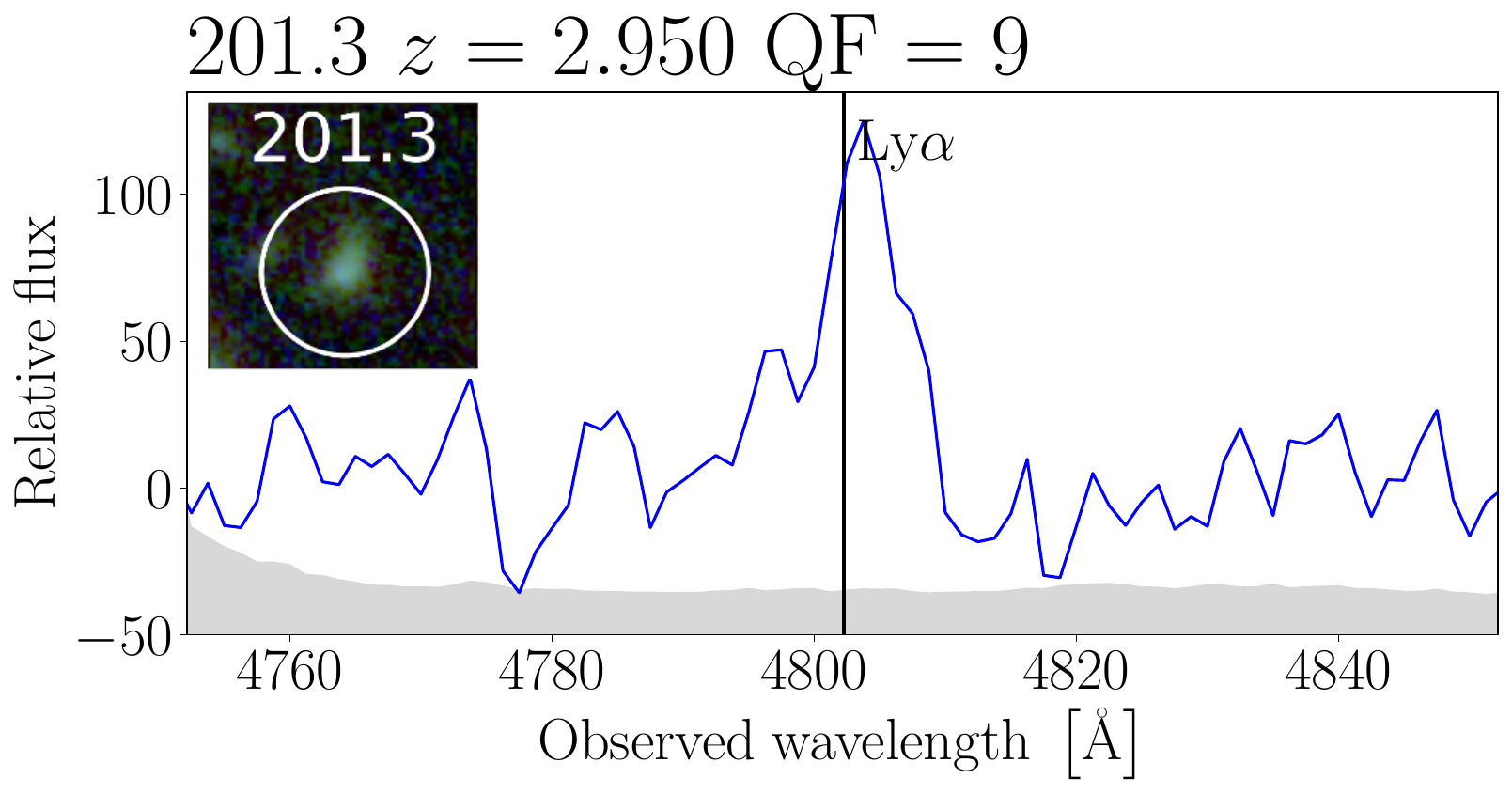}
  \includegraphics[width = 0.66\columnwidth]{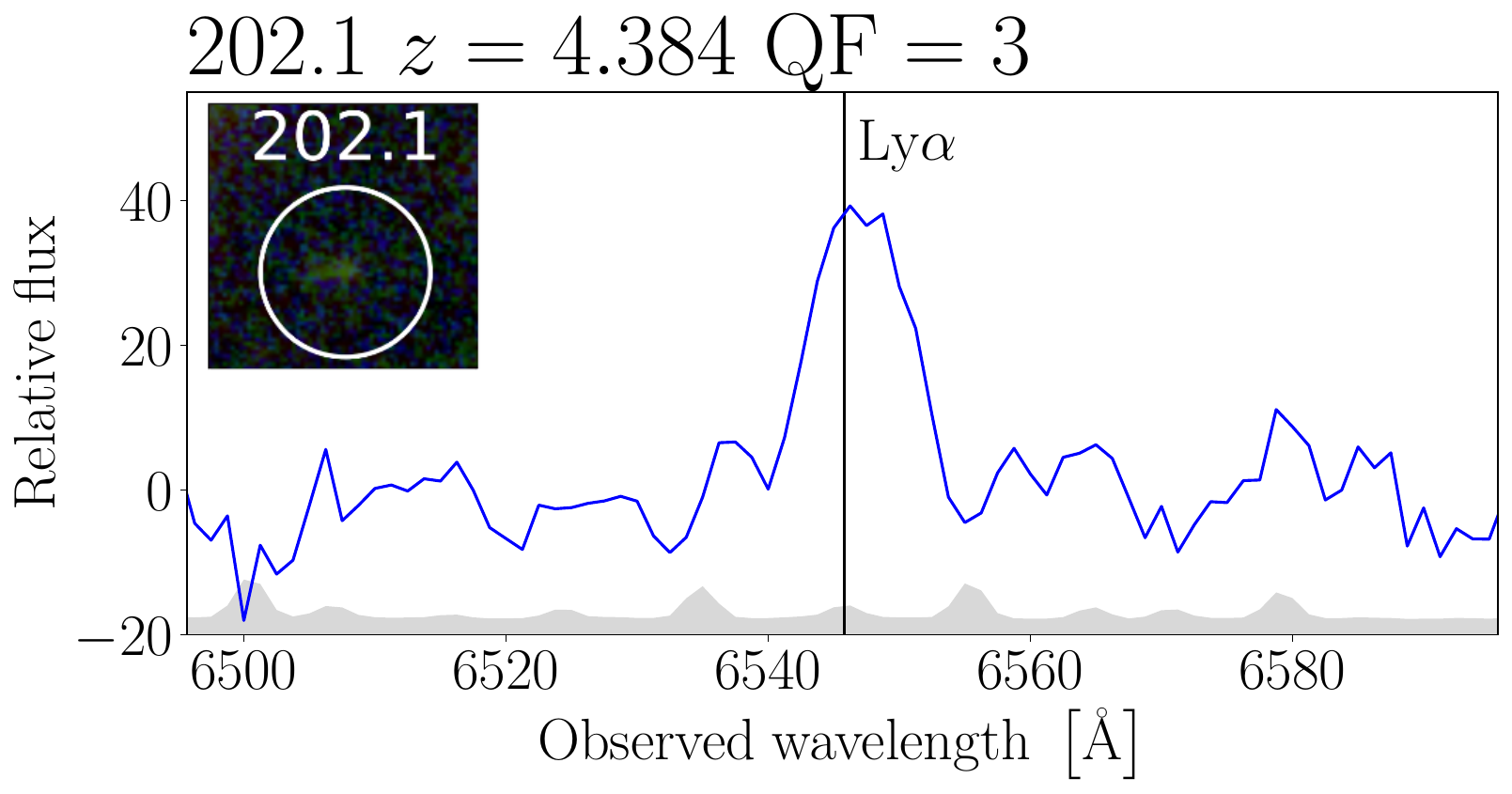}
  \includegraphics[width = 0.66\columnwidth]{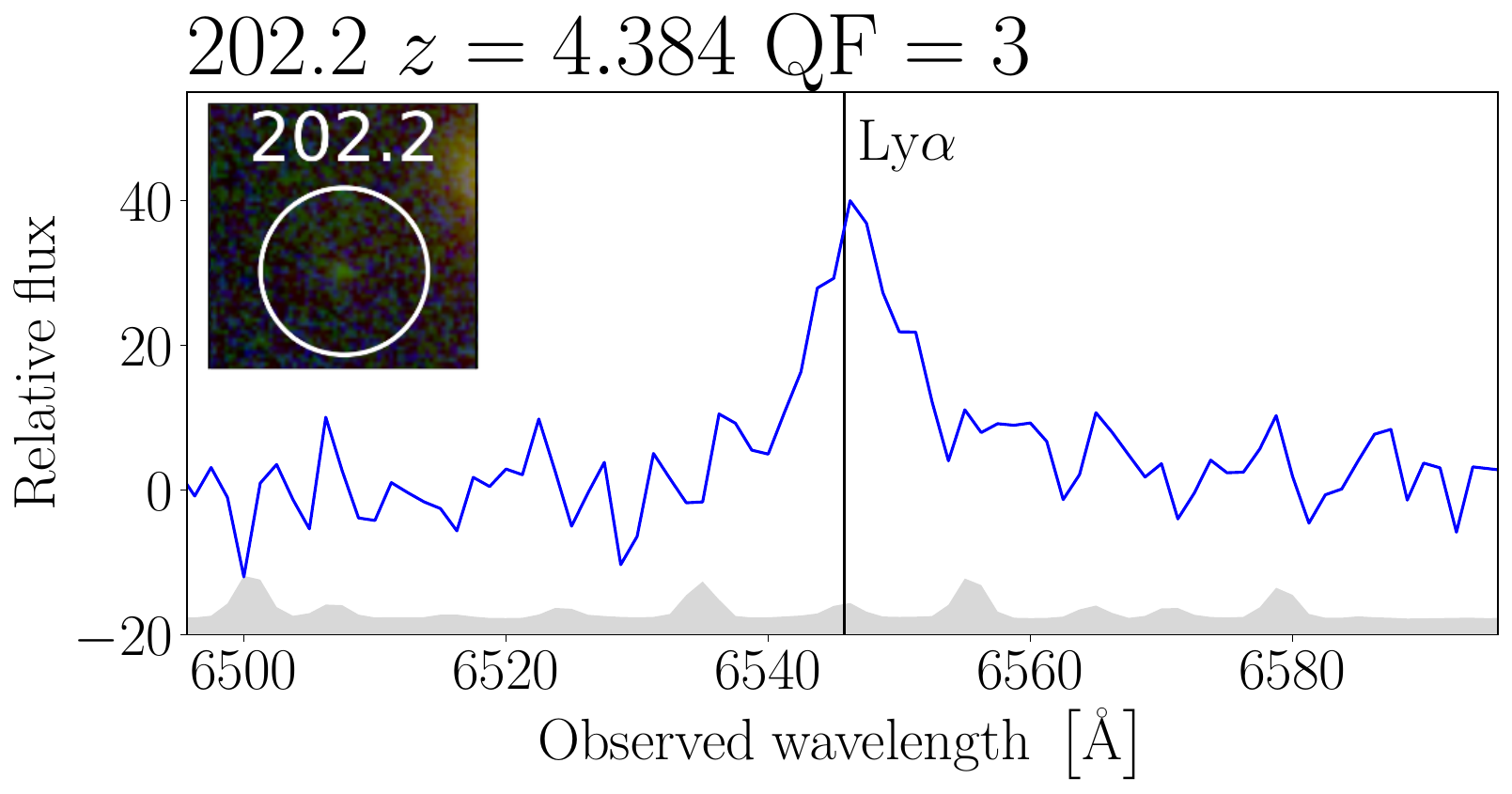}
  \includegraphics[width = 0.66\columnwidth]{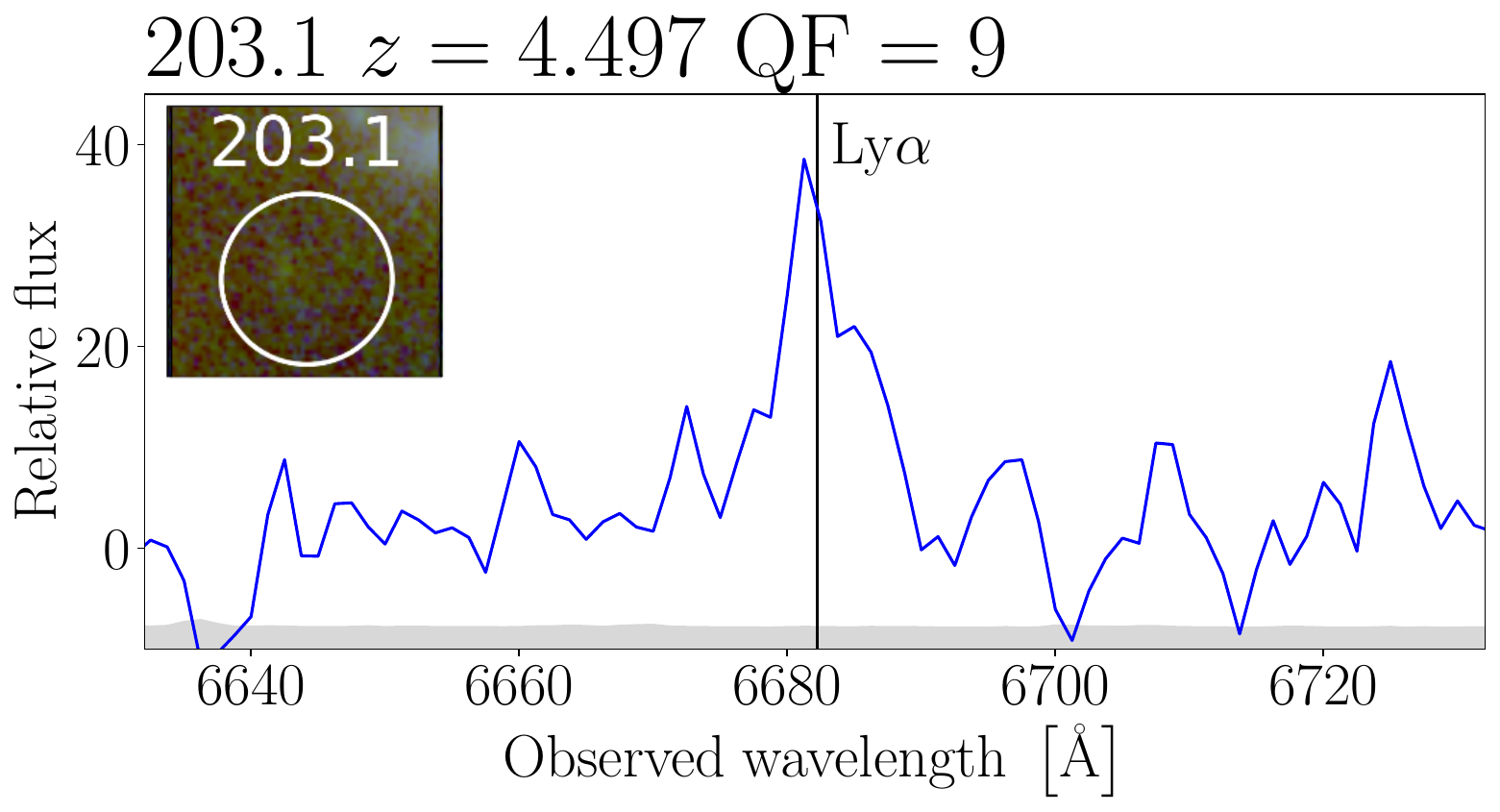}
  \includegraphics[width = 0.66\columnwidth]{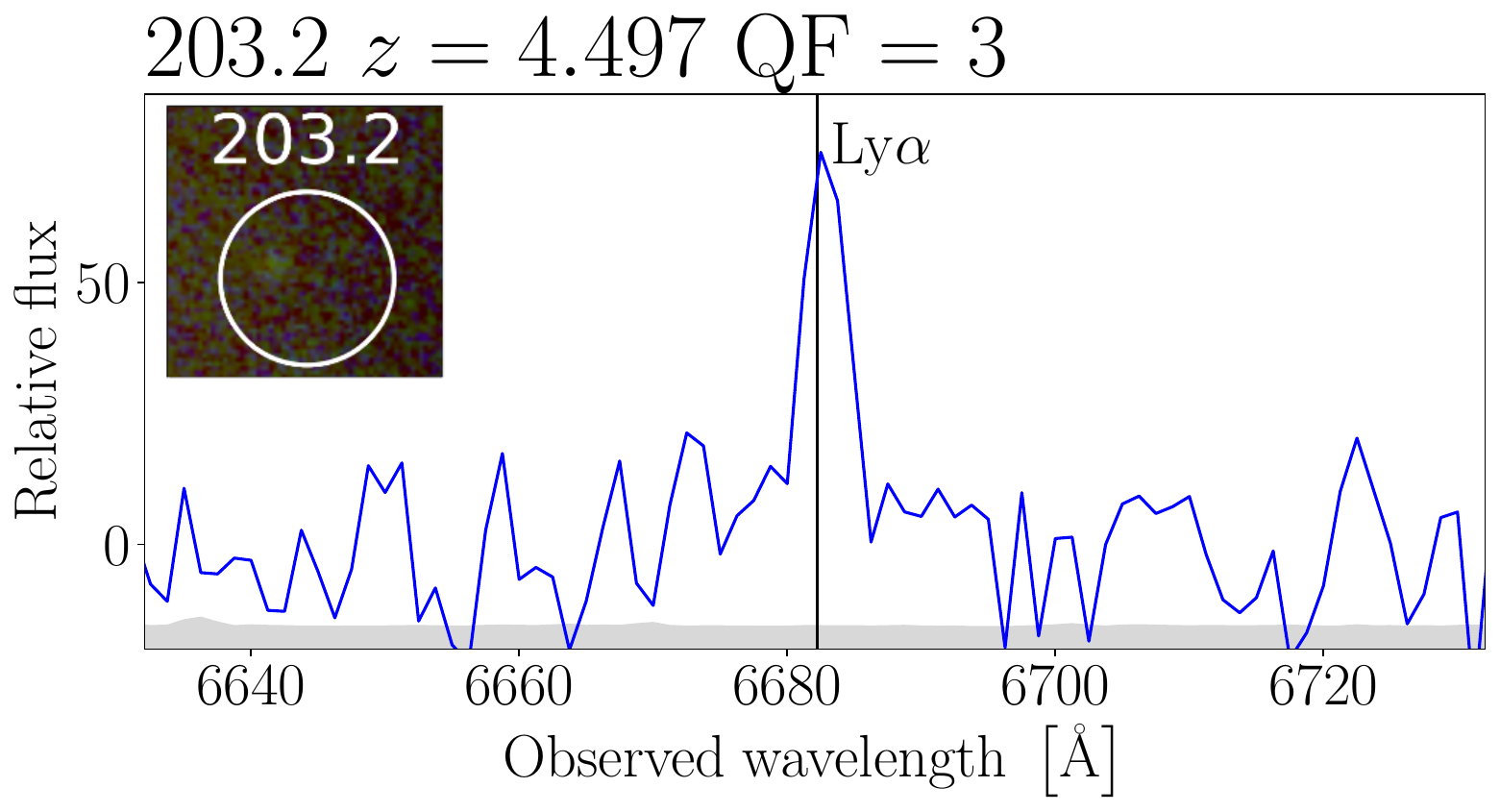}
  \includegraphics[width = 0.66\columnwidth]{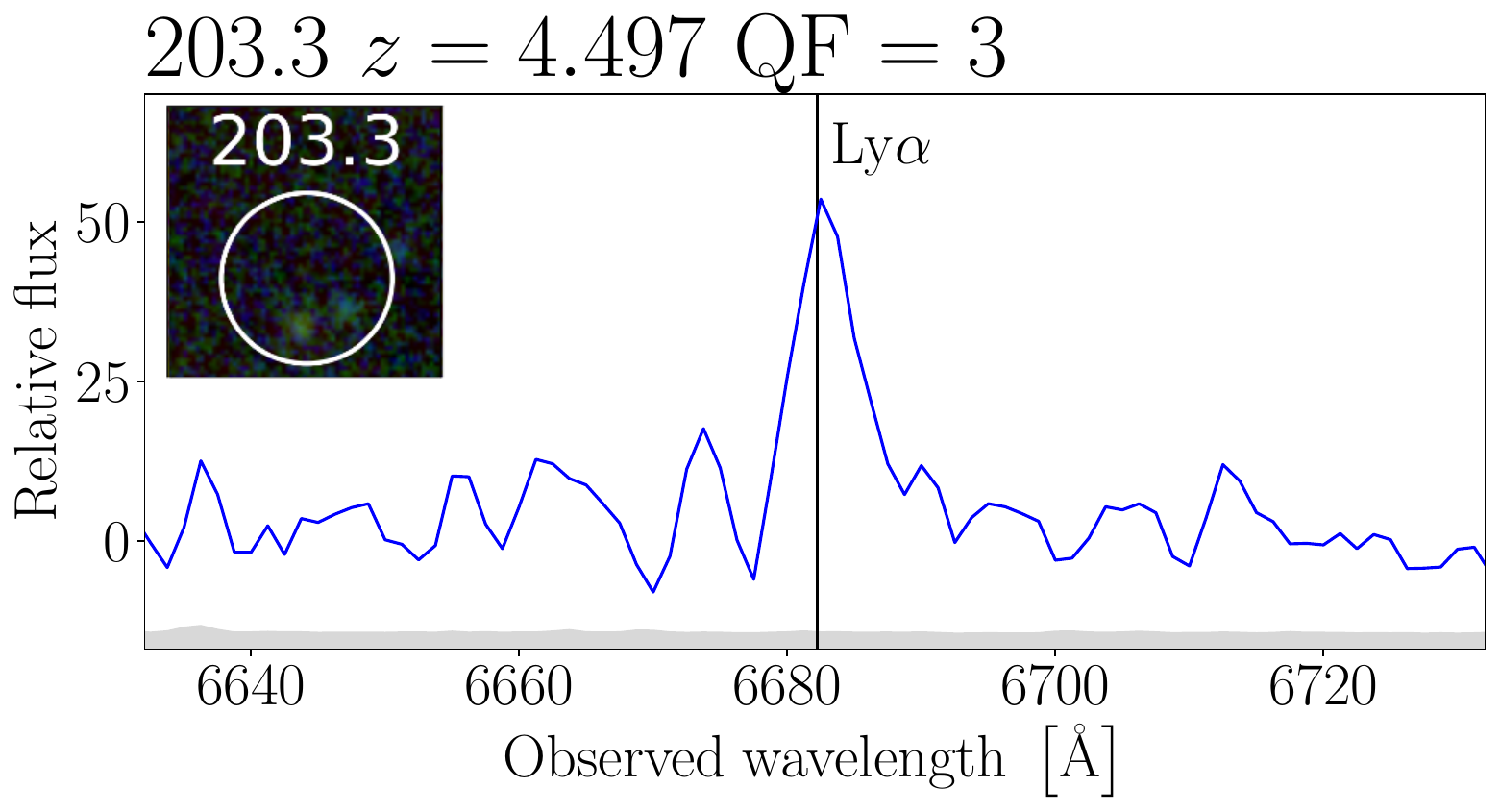}
  \includegraphics[width = 0.66\columnwidth]{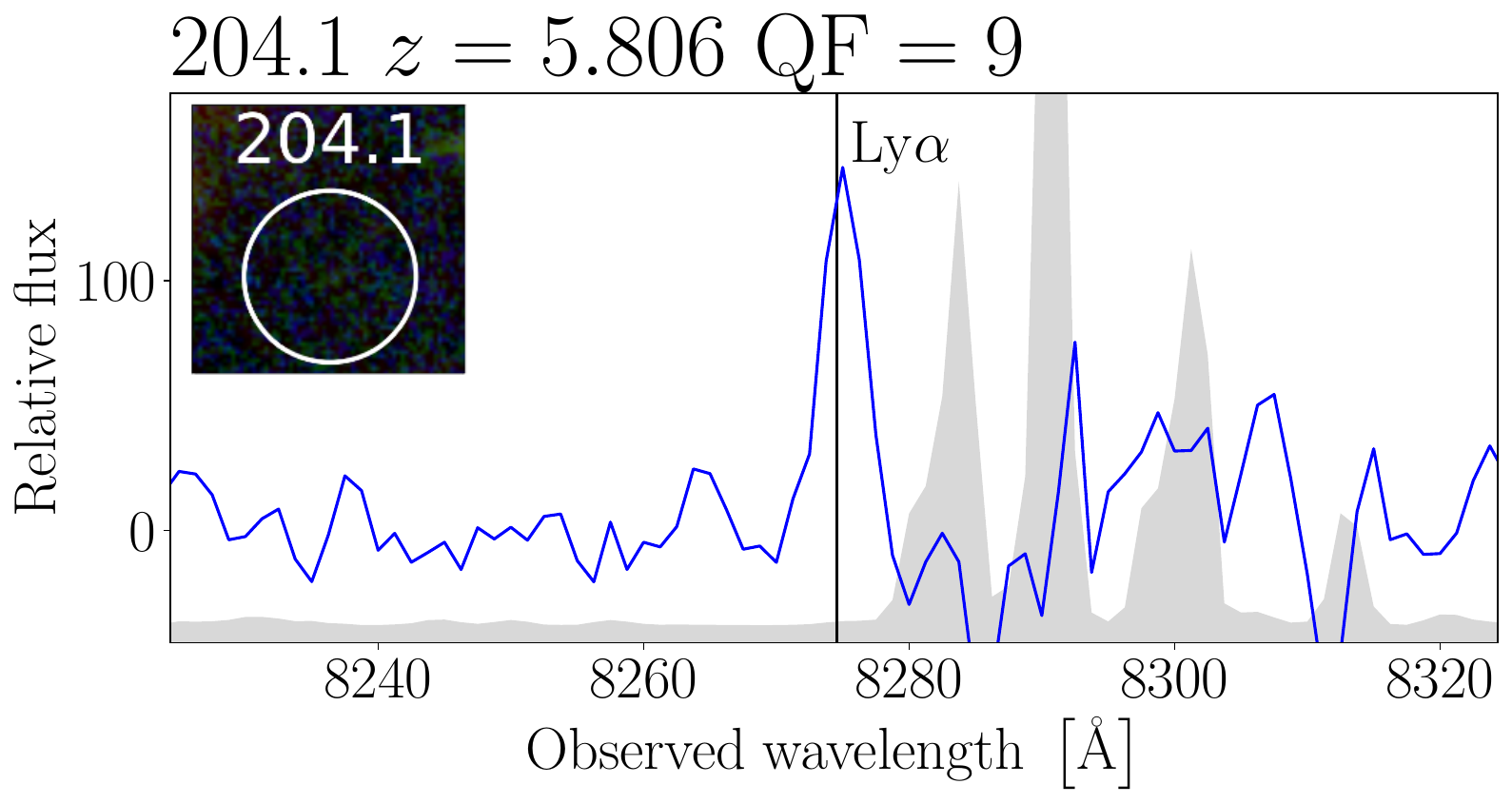}
  \includegraphics[width = 0.66\columnwidth]{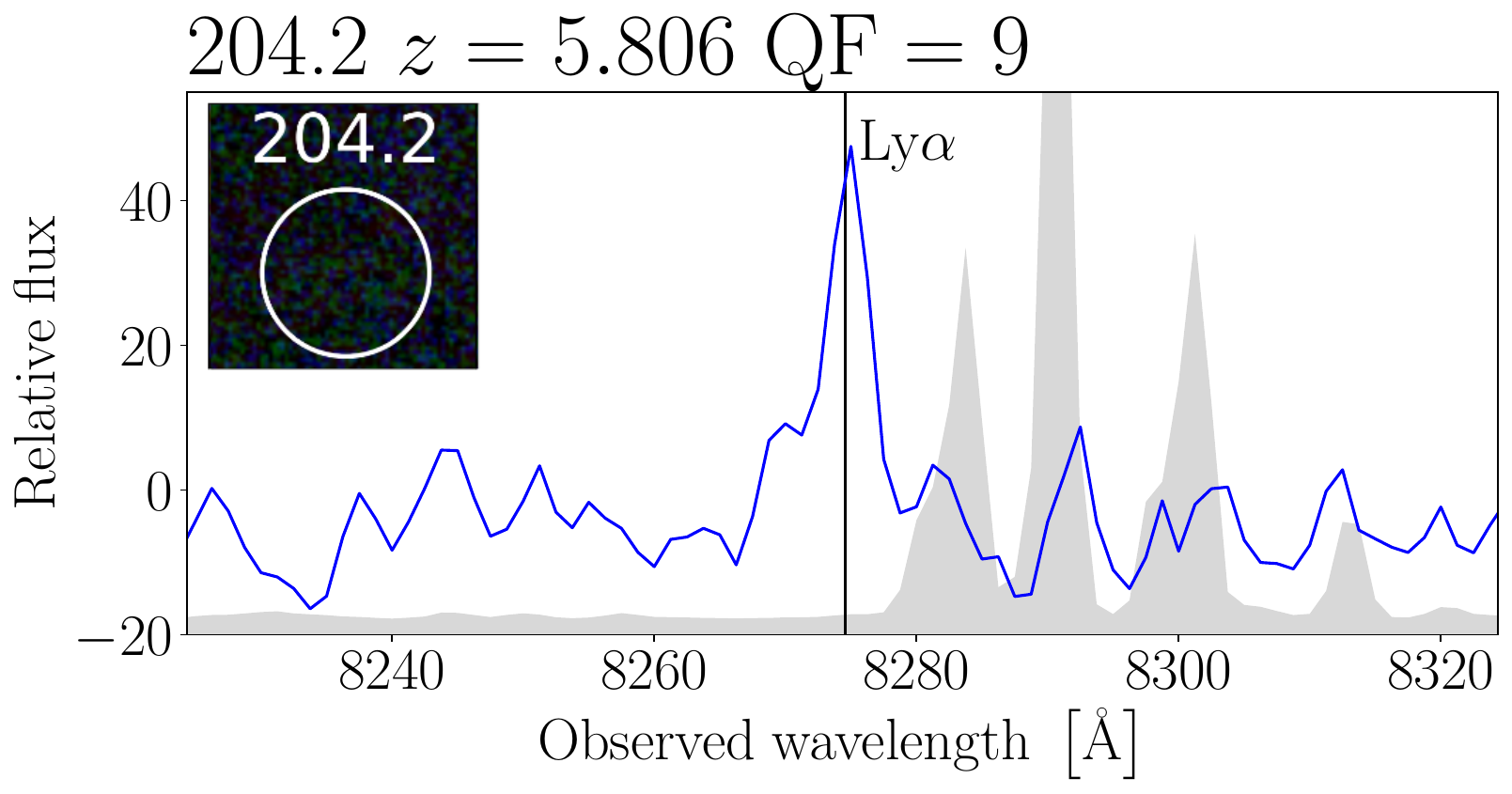}
  \includegraphics[width = 0.66\columnwidth]{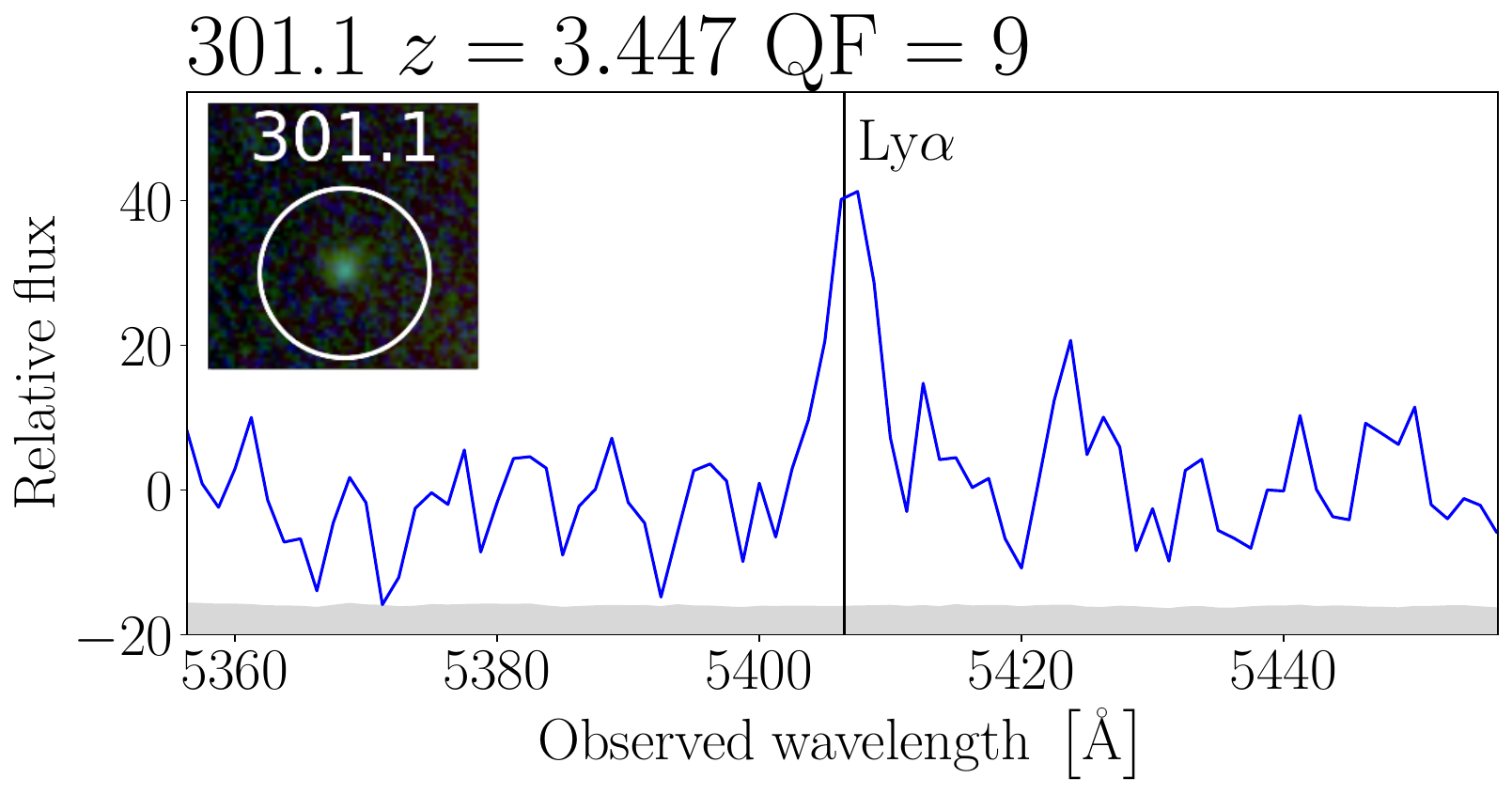}
  \includegraphics[width = 0.66\columnwidth]{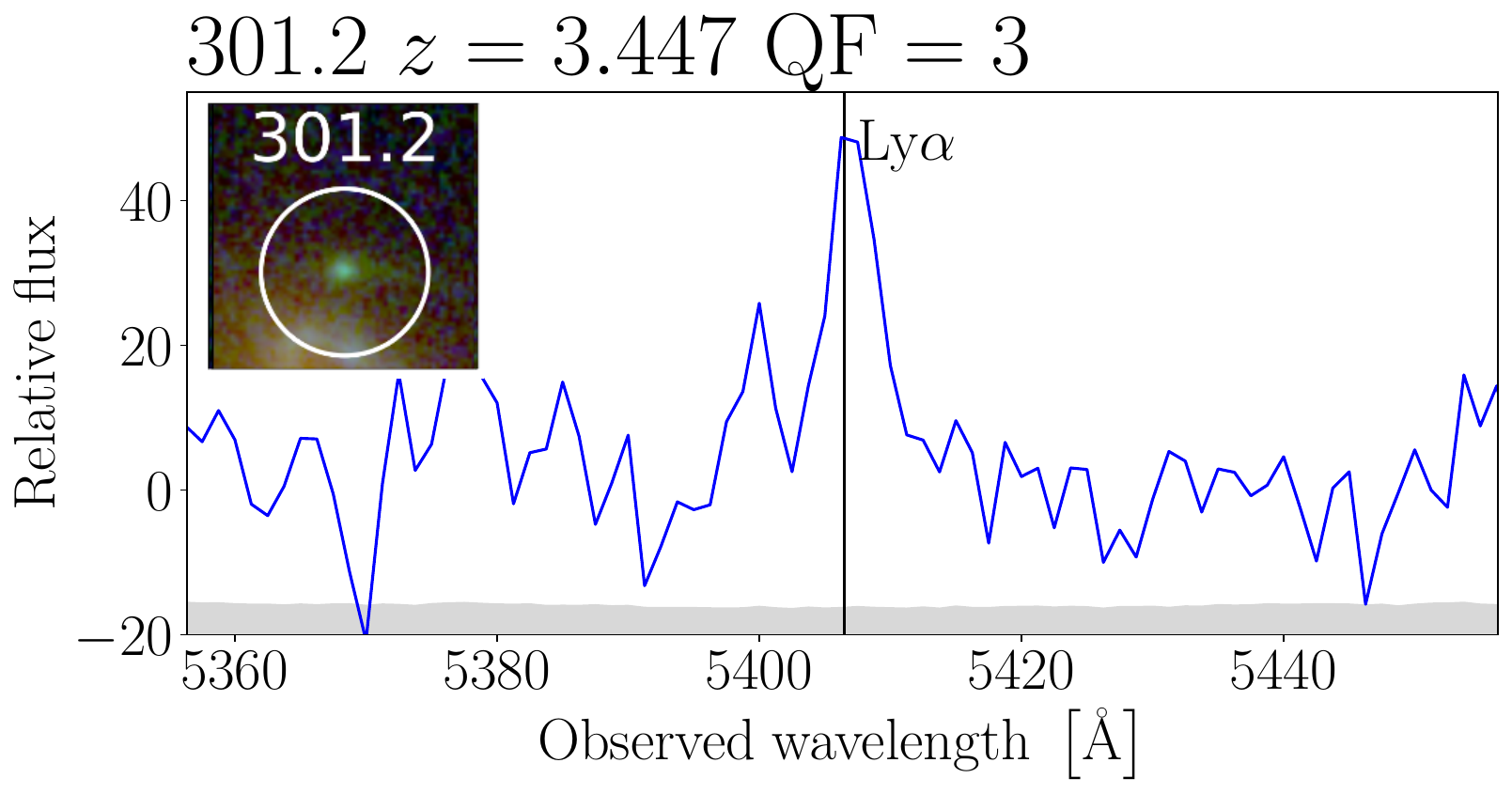}
  \includegraphics[width = 0.66\columnwidth]{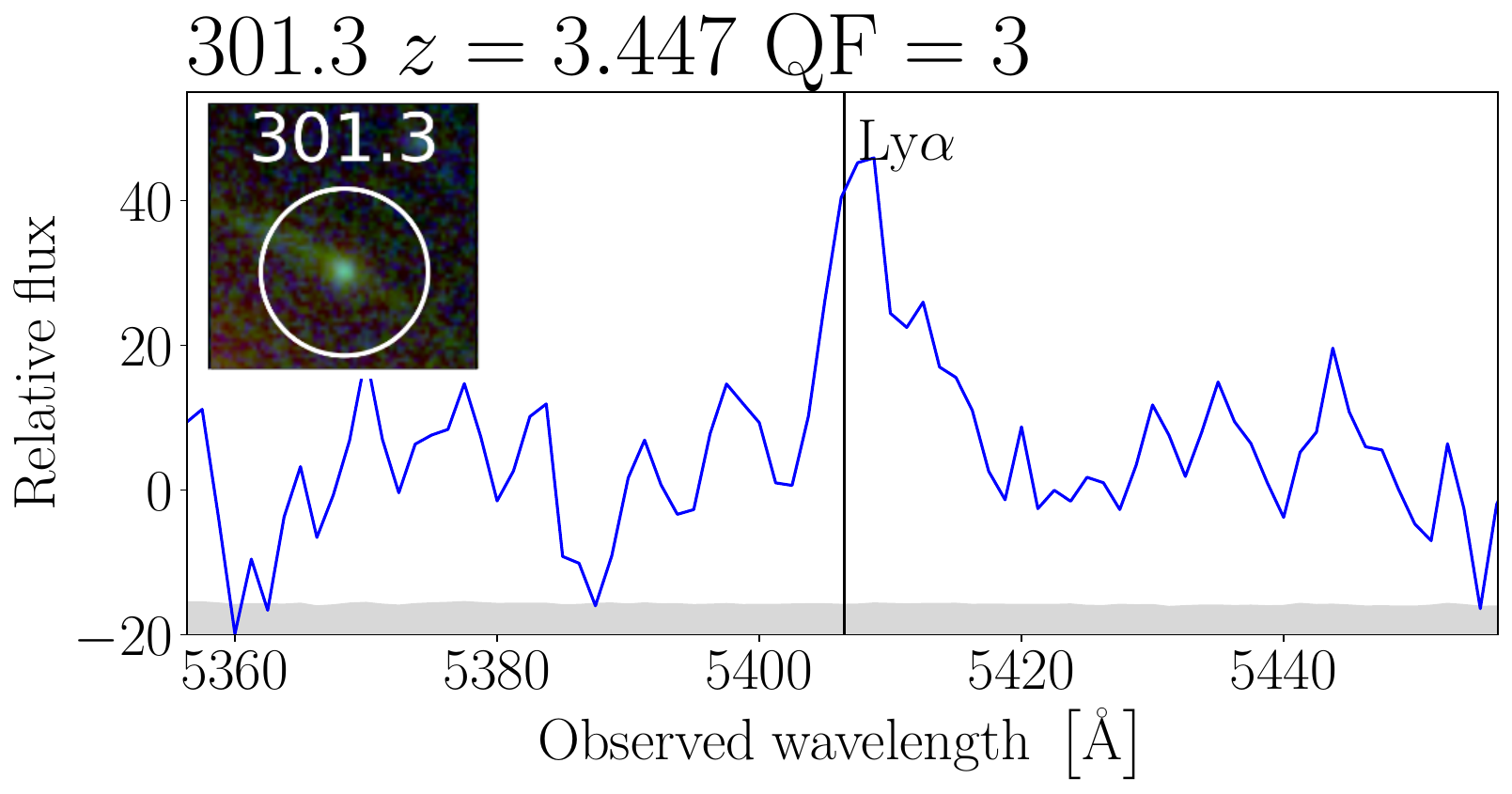}
  \includegraphics[width = 0.66\columnwidth]{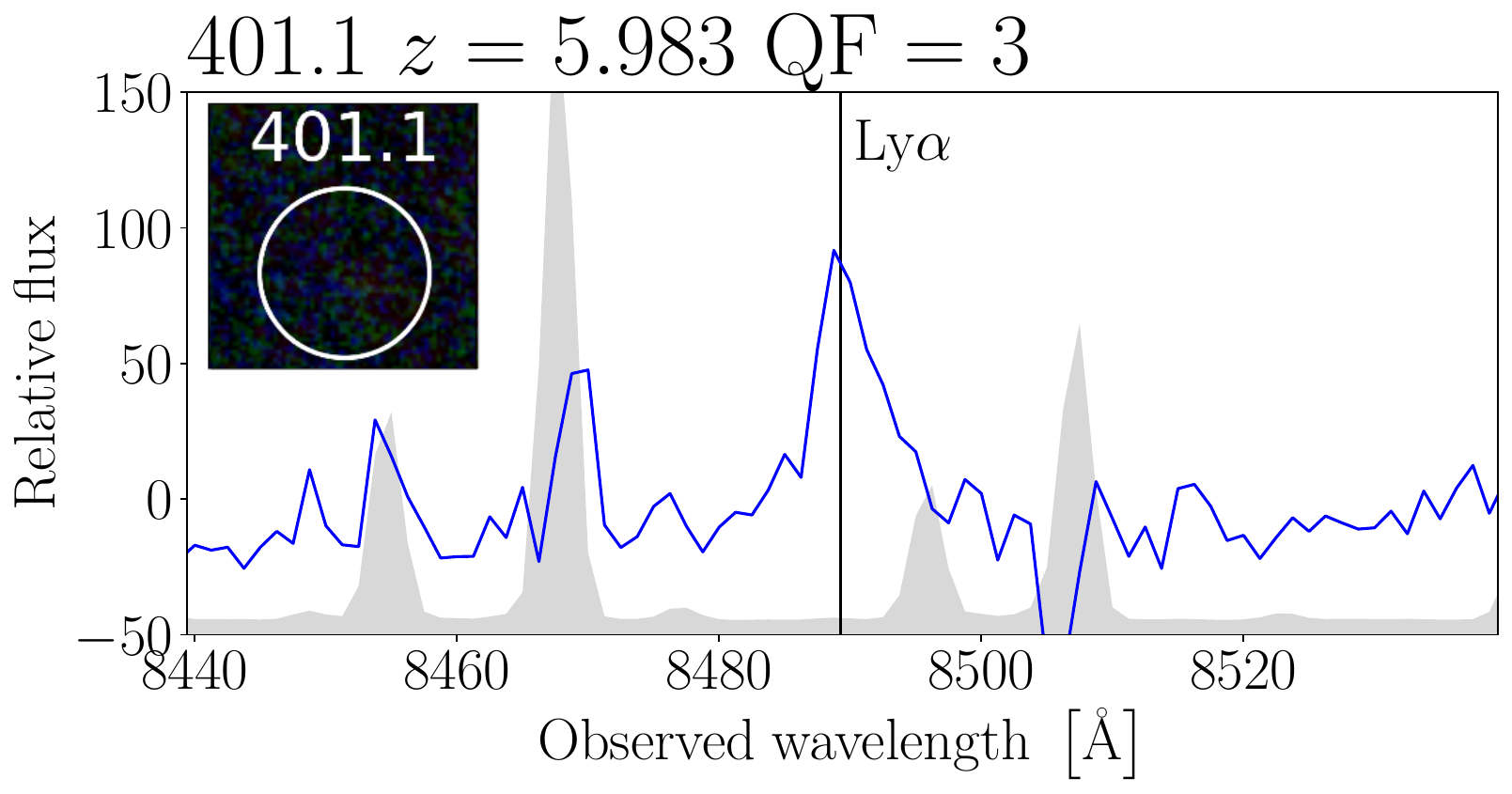}
  \includegraphics[width = 0.66\columnwidth]{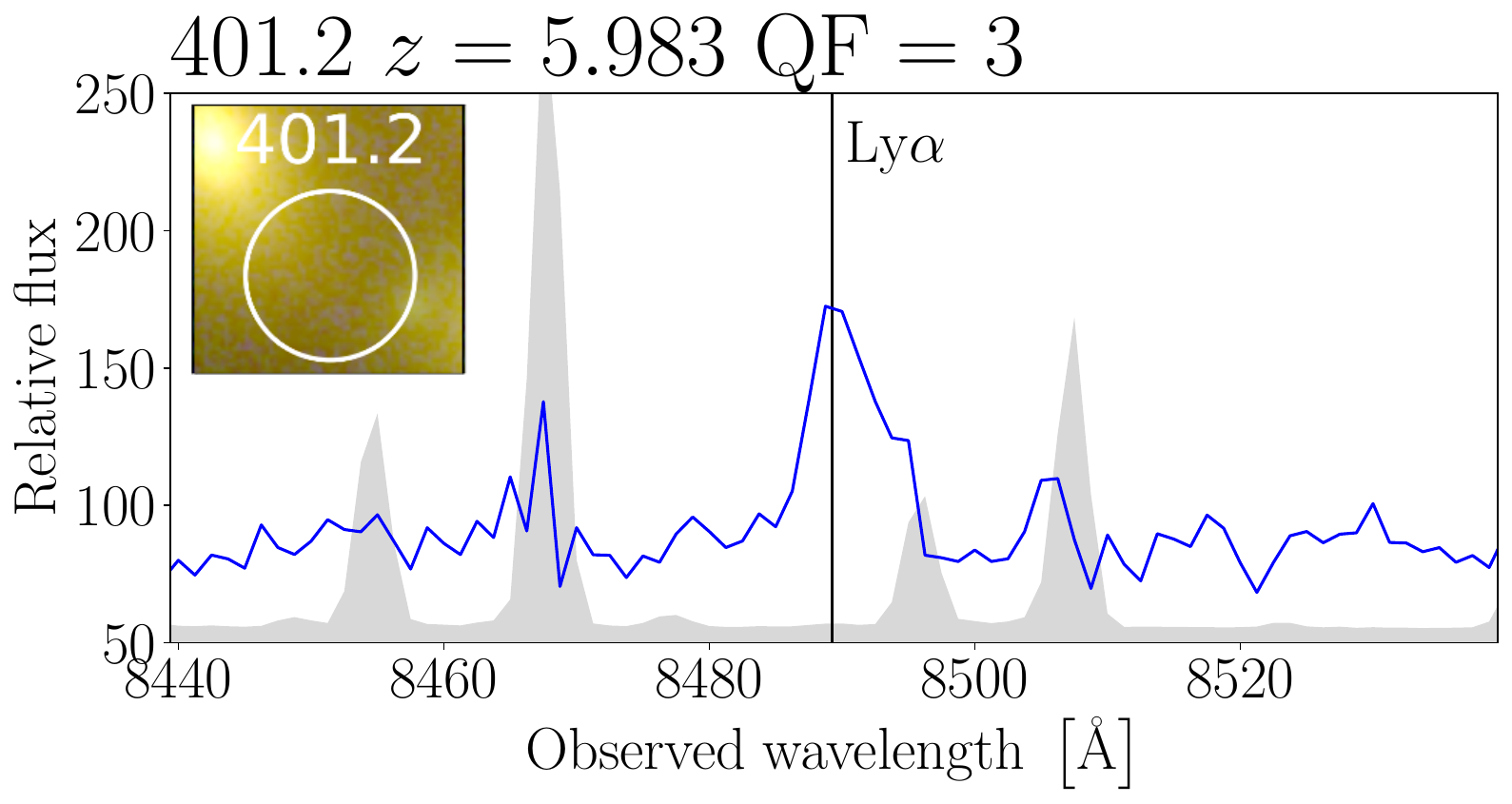}
  \includegraphics[width = 0.66\columnwidth]{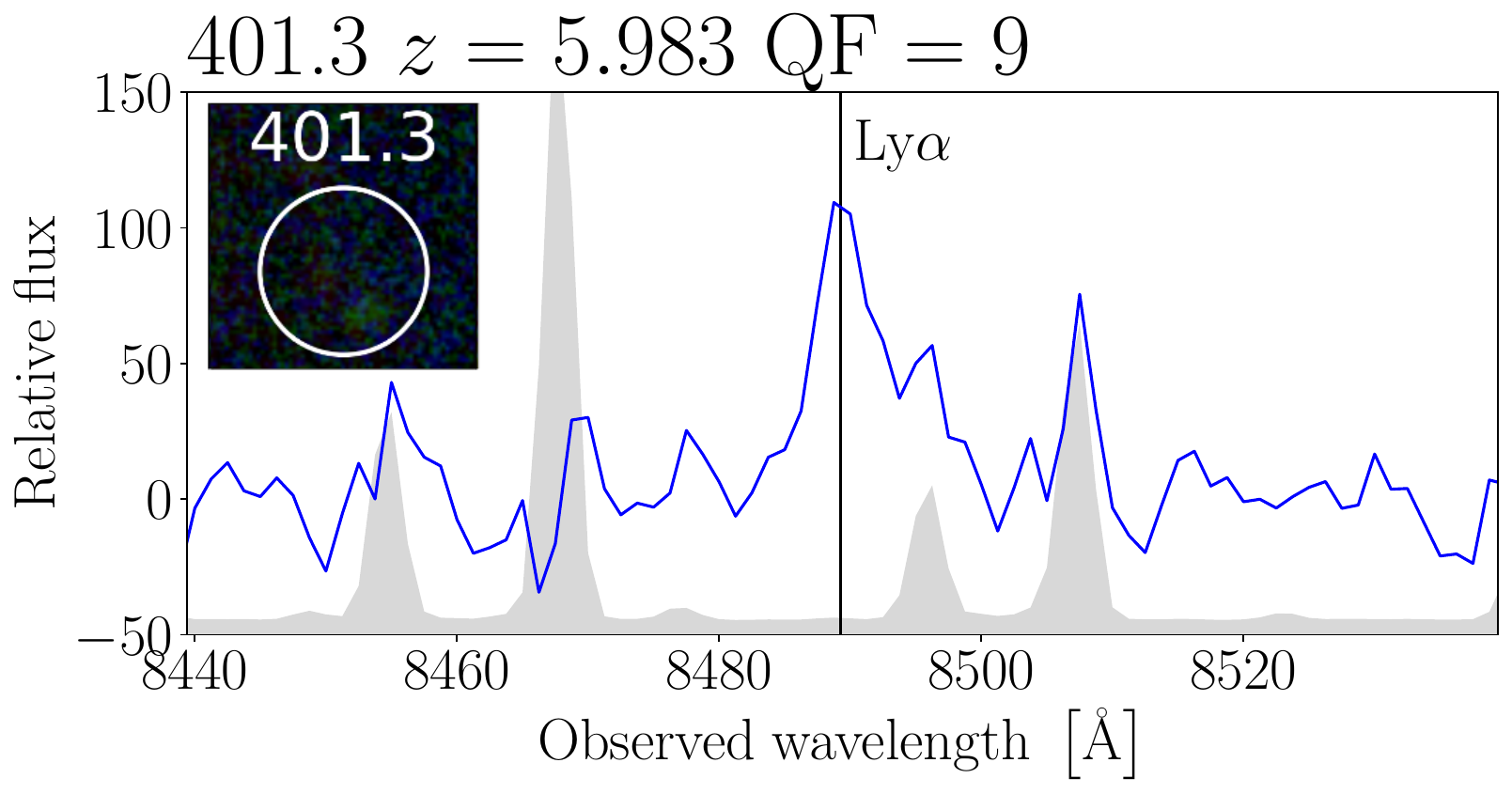}
  \includegraphics[width = 0.66\columnwidth]{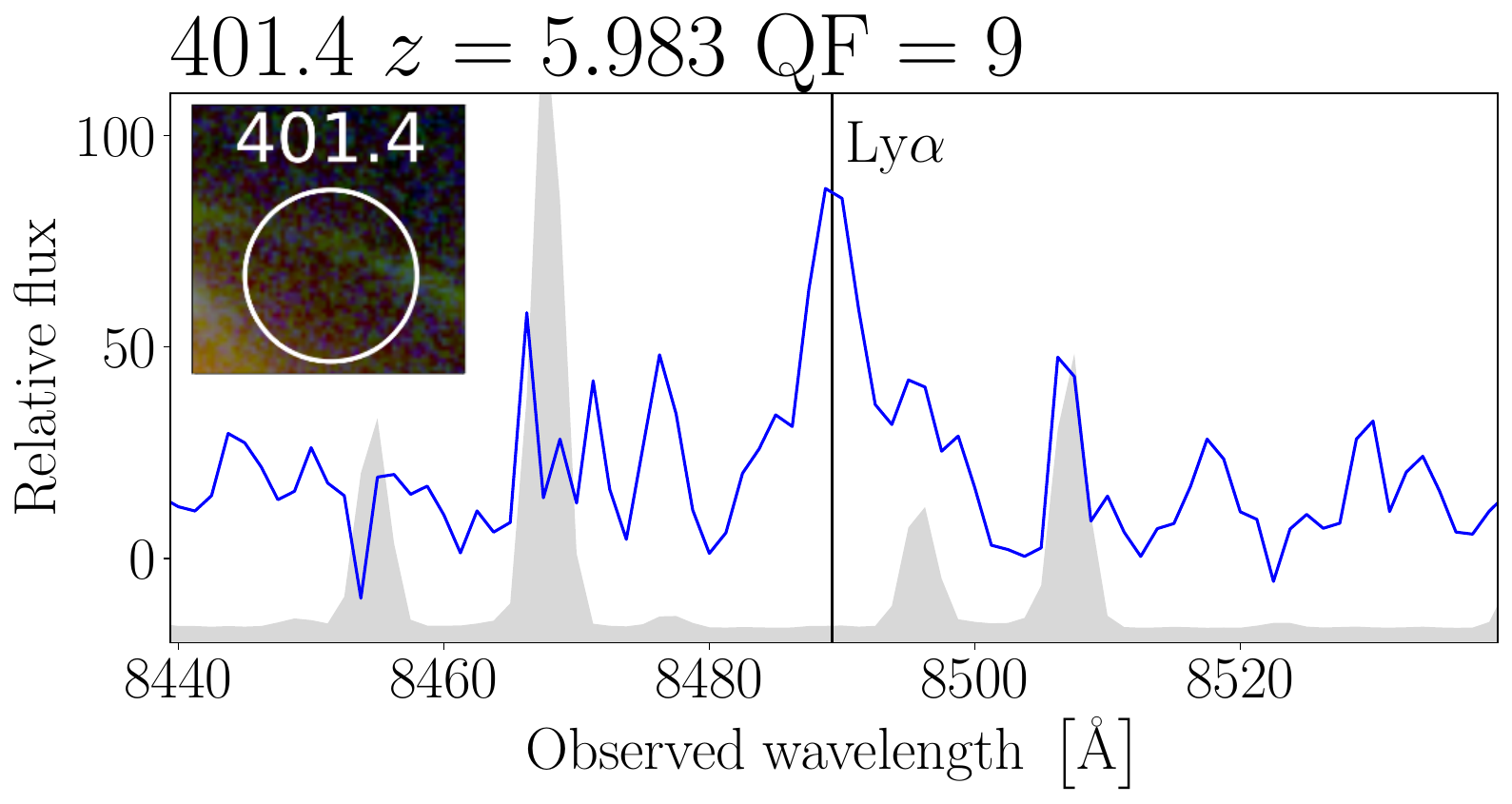}
  \caption{Spectra of the newly discovered multiple images. The vertical line indicates the observed position of the Lyman-$\alpha$ emission and the rescaled variance is shown in gray. ID, redshift and quality flag are indicated in each figure title. The \HST \, cutouts are $2\arcsec$ on each side, and north is up and east is left.}
  \label{fig:multiple_image_spectra}
\end{figure*}

Based on the newly obtained \MUSE \, redshifts (see Table~\ref{tab:z_muse}), we identified two sources that are multiply lensed by the galaxy group located in the northern region of MACS~1149 (see Fig.~\ref{fig:HFF_zoom}). One source (ID 301), lying at $z=3.447$, is triply lensed, and visible in the \HST \, images, while the other (ID 401), at $z=5.983$, is quadruply lensed and only detected in the \MUSE \, data. We show the corresponding spectra in Fig.~\ref{fig:multiple_image_spectra}. We further observe two sources (IDs 202 and 204), which are both doubly lensed. The first one is at $z=4.384$ and detected by \HST, while the other one, at $z=5.806$, was only identified through a single emission line in the \MUSE \, spectra (see  Fig.~\ref{fig:multiple_image_spectra}).

We carried out a dedicated search for simultaneously appearing spectral features when scanning the old and new, combined \MUSE \, data cube to detect possible multiple images. With this procedure, which joins an automatic detection and a visual inspection \citep[see Sect.~2.2 of][]{caminha23}, we found two multiply lensed sources (ID 201 and 203) that were previously unknown. Both are observed three times and, again, one (ID 203) of the two has only been detected by \MUSE. We show their spectra in Fig.~\ref{fig:multiple_image_spectra} as well. 

We summarize all the newly identified multiple images in Table~\ref{tab:sources}, which are also marked in Fig.~\ref{fig:HFF} (white cross). They are included in our lensing model presented in Sect.~\ref{sec:modeling}, complemented by those published by \citetalias{grillo16} and the fifth image SX of SN Refsdal, published by \citet[][marked in white in Fig.~\ref{fig:HFF}]{kelly16b}. 

\begin{table}[t!]
    \caption{New multiple images, in addition to those from Tables 2 and 3 of \citetalias{grillo16}, used in our strong lensing model of MACS~1149.}
    \begin{center}
    \begin{tabular}{cccccc}
    \hline \hline \noalign{\smallskip}
ID   &RA             &Dec.           &$z_\text{sp}$ &QF &identification\\
& (J2000) & (J2000) & & & \\
\noalign{\smallskip} \hline \noalign{\smallskip}
SX$^a$&177.40021 &22.39681 &1.489 &3  & \HST \\
201.1 &177.40562 &22.40243 &2.950 &3  & \HST \\
201.2 &177.40216 &22.39675 &2.950 &9  & \HST \\
201.3 &177.39530 &22.39182 &2.950 &9  & \HST \\
202.1 &177.40014 &22.40415 &4.384 &3  & \HST \\
202.2 &177.39834 &22.40378 &4.384 &3  & \HST \\
203.1 &177.40028 &22.39641 &4.497 &9  & MUSE \\
203.2 &177.39575 &22.40000 &4.497 &3  & MUSE \\
203.3 &177.39457 &22.39442 &4.497 &3  & MUSE \\
204.1 &177.40043 &22.40389 &5.806 &9  & MUSE \\
204.2 &177.39949 &22.40376 &5.806 &9  & MUSE \\
301.1 &177.39504 &22.41269 &3.447 &9  & \HST \\
301.2 &177.39353 &22.41307 &3.447 &3  & \HST \\
301.3 &177.39285 &22.41287 &3.447 &3  & \HST \\
401.1 &177.39591 &22.41239 &5.983 &3  & MUSE \\
401.2 &177.39357 &22.41142 &5.983 &3  & MUSE \\
401.3 &177.39168 &22.41153 &5.983 &9  & MUSE \\
401.4 &177.39319 &22.41295 &5.983 &9  & MUSE \\
    \noalign{\smallskip} \hline 
    \end{tabular}
    \end{center}
\textbf{Notes.} Columns give, from left to right, the identification number ID, right ascension, declination, spectroscopic redshift $z_\text{sp}$, the corresponding redshift uncertainty in terms of a quality flag QF, and whether the image position is measured in the \HST\ images or only from the \MUSE \, data cube. Considered quality flags are: 3 very secure ($\delta z <$ 0.001) and 9 secure (based on a single emission line).\\
$^a$ Published in \citet{kelly16b}.
    \label{tab:sources}
\end{table}

\section{Lens modeling}
\label{sec:modeling}

The strong lensing analysis follows closely that presented in \citetalias{grillo16}, so we refer to that work, as well as to \citet{grillo15}, for a detailed discussion of the modeling and statistical analysis. In the following, we summarize the procedure and highlight the differences compared to the previous model from \citetalias{grillo16}. 

The cluster total mass model was obtained with the Gravitational Lens Efficient Explorer \citep[\GLEE,][]{suyu10a_GLEE, suyu12b_GLEE} software. This software has been widely used for strong-lens modeling, in particular for time-delay cosmography on galaxy-scale \citep[e.g.,][]{wong20, ertl23, shajib22} as well as on cluster-scale \citep{grillo18, grillo20, grillo24} systems. The software supports mainly simply parameterized and physically motivated mass density profiles, as introduced below, to describe the total mass distribution of a lens. It provides several optimization and Bayesian sampling algorithms to infer the best-fitting parameter values and to sample their parameter space. Here, we preferentially relied on the simulated annealing, the Markov chain Monte Carlo (MCMC) technique, and its parallelized version {\tt emcee} \citep{foreman-mackey13}, with the main advantage of decreasing the computational time. We further used and extended \textit{glee\_tool.py} \citep{schuldt23b}, a code that supports several tools for modeling with \GLEE \, and helps to reduce the user input time.

In the following, we detail the adopted mass density profiles, how we include the multiple images as constraints, and which model configurations we consider. Finally, we give a summary of our final model and the corresponding results.

\subsection{Multiple image systems}
\label{sec:modeling:sources}

The total mass distribution of MACS~1149 is reconstructed from securely identified multiply lensed sources. We include 62 multiple images that belong to 18 different knots of the SN Refsdal host galaxy (see Table~3 of \citetalias{grillo16}), and 26 multiple images from ten different sources (see Table~2 in \citetalias{grillo16}). Moreover, we now added the fifth image of SN Refsdal (referenced as SX), which has been detected in December 2015 \citep{kelly15_ATel, kelly16b}. These previously known systems have now been complemented thanks to our recent \MUSE \, observations (see Sect.~\ref{sec:muse}), with 17 new multiple images from six different sources, listed in  Table~\ref{tab:sources}. The combination resulted in a total of 106 multiple images from 34 different families covering a redshift range between 1.24 and 5.98, with most constraints at $z_\text{SN}=1.489$ from SN Refsdal and its host. The redshifts of all multiply lensed sources are highlighted with orange lines in Fig.~\ref{fig:redshifts}.

While 31 families are spectroscopically confirmed, three are photometrically identified systems from the ``gold'' sample presented in Table~3 of \citet{treu16}. As in previous strong lensing models, these last systems are included with a redshift value free to vary. We adopted for them a broad range between zero and six, with a flat prior. This leads to three free parameters, in addition to their variable source positions ($34\times2=68$ free parameters), beside the mass model free parameters (see Sect.~\ref{sec:modeling:cm} and Sect.~\ref{sec:modeling:halo} for more details), while providing $(106\times 2=)~212$ positional constraints for the model.

\subsection{Cluster member mass component}
\label{sec:modeling:cm}

We included in our total mass model all the 308 cluster members identified in Sect.~\ref{sec:muse:analysis:members} with the same total mass density profile. These mass components are centered on the galaxy luminosity centroid positions (see Table~\ref{tab:clustermembers}) and described by a dual pseudo-isothermal elliptical mass distribution \citep[dPIE;][]{eliasdottir07, suyu10a_GLEE} with axis ratio $q\equiv 1$ and vanishing core radius, such that the convergence (or dimensionless surface mass density) $\kappa$ of the cluster member $j$ can be expressed as
\be
\kappa_j(x,y) = \frac{\theta_{\text{E},j}}{2} \left( \frac{1}{\sqrt{x^2+y^2}} - \frac{1}{\sqrt{x^2+y^2+r_{\text{t},j}^2}} \right) \, .
\ee
The values of the Einstein radius $\theta_{\text{E},j}$ and of the truncation radius $r_{\text{t},j}$ are connected to those of the brightest cluster galaxy (BCG, ID 166 in Table~\ref{tab:clustermembers}) through their measured total luminosity $L_j$ or magnitudes $m_j$ (\HST \, F160W band, see Table~\ref{tab:clustermembers}) following the scaling relations
\be
\theta_{\text{E,}j} = \theta_\text{E,BCG} \left(\frac{L_j}{L_\text{BCG}}\right)^{0.7}  = \theta_\text{E,BCG} \times 10^{0.28(m_\text{BCG}-m_j)}
\ee
and 
\be 
r_{\text{t,}j} = r_\text{t,BCG} \left(\frac{L_j}{L_\text{BCG}}\right)^{0.5} = r_\text{t,BCG} \times 10^{0.2(m_\text{BCG}-m_j)} \, ,
\ee
respectively. This means that we are able to effectively vary only two parameters to determine the total mass distribution of all cluster members, namely the Einstein radius $\theta_\text{E,BCG}$, and the truncation radius $r_\text{t,BCG}$ of the BCG.

Since the region around SN Refsdal is particularly interesting for cosmological applications \citep[see e.g.,][]{grillo20, grillo24} and very well constrained through 62 multiple images, associated with 18 knots of the SN host galaxy (see Sect.~\ref{sec:muse:analysis:images} and Table~3 in \citetalias{grillo16}), we treated two galaxy members separately. These two members are located close the images of SN Refsdal, and we allowed, as done in \citetalias{grillo16}, their dPIE profiles to have the Einstein and truncation radii free to vary outside of the scaling relations defined above. We further introduced an ellipticity term. This added up to eight additional free parameters, which is a small number, given the high number of observables in that region, and which are necessary to reconstruct the Fermat potential around the multiple images of SN Refsdal properly.

\subsection{Cluster-scale mass components}
\label{sec:modeling:halo}

In addition to including a total mass component for each cluster member, we also modeled the cluster mass distribution on larger scales. This is mainly the dark matter (DM) of the cluster halos not associated with the individual galaxies, such that we refer to them as DM halos in the following. This mass term is described by multiple pseudo-isothermal elliptical mass distributions \citep[PIEMDs][]{kassiola93} characterized by six parameters: the coordinates of the center $x_\text{H}$ and $y_\text{H}$; the axis ratio $q_\text{H}$; its orientation $\phi_\text{H}$; the halo strength (or Einstein radius) $\theta_\text{E,H}$; and the core radius $r_\text{c,H}$. 

Because of the complex merging state of MACS~1149, as revealed by Chandra X-ray observations \citep[see][]{ogrean16, golovich16}, we followed the results of previous strong-lensing analyses \citep[e.g.,][]{smith09, rau14, grillo16} and included in the initial cluster total mass model three PIEMD profiles. This results in 18 free parameters for the cluster-scale components. 

\subsection{Mass models}
\label{sec:modeling:tests}

We started with the parametrization of \citetalias{grillo16}, as described above, which means we include 308 dPIE profiles for the cluster members and three DM halos. By incorporating the new multiple images listed in Table~\ref{tab:sources} and adopting the usual $\chi^2$ definition, the model cannot reproduce the correct image multiplicity of some families. This holds particularly for the two families around the galaxy group in the north (IDs 301 and 401, see Fig.~\ref{fig:HFF_zoom}). This might be due to an oversimplified parametrization of the total lens mass distribution, as we included only the mass of the cluster members through the scaling relations, while fully neglecting background and foreground galaxies and the specific DM halo associated with this group. 

To avoid models with wrong image multiplicity, we modify the usual $\chi^2$ definition in the following way
\be
\chi^2 = \sum_{j=1}^{N^\text{fam}} \left\{ \begin{array}{cc}
     \sum_{k=1}^{N_j^\text{obs}}  \frac{\left(x_{j,k}^\text{obs} - x_{j,k}^\text{pred} \right)^2 }{\Delta x_{j,k}^2} + \frac{\left(y_{j,k}^\text{obs} - y_{j,k}^\text{pred} \right)^2 }{\Delta y_{j,k}^2} &  \text{if~} N_j^\text{obs} \leq N_j^\text{pred}\\
     \infty & \text{otherwise}
     \label{eq:chi2}
\end{array} \right. ,
\ee
with the number of image families $N^\text{fam}$ (in our case 34), the observed and predicted number of multiple images that belong to family $j$ as $N^\text{obs}_j$ and $N^\text{pred}_j$, respectively. The observed position of image $k$ from family $j$ is located at ($x_{j,k}^\text{obs}$, $y_{j,k}^\text{obs}$) and matched to the closest model-predicted image position at ($x_{j,k}^\text{pred}$, $y_{j,k}^\text{pred}$), while a predicted image cannot be matched more than once to an observed image. The latter requirement is often neglected and can lead to a lower $\chi^2$ value. Although this definition introduces artificial steep walls in the $\chi^2$ function, we did not find any problems with this simple model rejection. Indeed, it should be possible to find the global minimum, particularly when using simulated annealing starting from high temperatures. In addition, once a parameter set that predicts enough multiple images for all different families was found, the MCMC walkers then explore the parameter space around those parameter values which are expected to lead to models with enough predicted images. In that case, the $\chi^2$ definition reduces to the usual $\chi^2$ function, without steep walls to reject specific models. Another option was recently introduced by \citet{ertl24} for the galaxy-scale lensing system HE0230$-$2130. They checked a posteriori the MCMC chains for the correct image multiplicity. This allowed \citet{ertl24} to gain insights into the preferred total mass parametrization of HE0230$-$2130. We note that this procedure is computationally significantly more expensive, as it also samples models predicting too few multiple images.

We further note that this modified definition still allows more model-predicted images than are included as observables in the model. These additional images could be, for instance, undetected due to a low magnification value, outside of the field of view, or that their exact position has not yet been confirmed. The image position uncertainties are described by $\Delta x_{j,k}$ and $\Delta y_{j,k}$, respectively, in $x$ and $y$ direction. As a consequence, we also modified the usual RMS statistics, and defined it in an analogous way to the expression in Eq.~\ref{eq:chi2}, resulting in a RMS value equal to or higher than that obtained with the usual definition. With these additional criteria, we were able to build models with the initial parametrization from \citetalias{grillo16} and correct image multiplicity, but with notably higher $\chi^2$ and RMS values. 

Throughout our modeling, we adopted a constant positional uncertainty $\sigma = \Delta x = \Delta y$ that is equal for all images identified in the HFF data, and doubled that value for the images detected only in the \MUSE \, data (i.e., $\sigma = 2 \times \Delta x = 2 \times \Delta y$, see Table~\ref{tab:sources}) as commonly done \citep[e.g.,][]{caminha17, bergamini19, bergamini21}. The larger uncertainty associated to the images detected only in \MUSE \, can be ascribed to the lower \MUSE \, spatial resolution and to the possible small offsets relative to the \HST \, images (see Sect.~\ref{sec:muse:reduction} for more details on the alignment). This uncertainty, denoted as $\sigma$ in Fig.~\ref{fig:rms}, was finally adjusted to $\Delta x = \Delta y  = 0.32\arcsec$, to obtain a $\chi^2$ value similar to the number of degrees of freedom (i.e., to get a reduced $\chi^2$ value of about 1). This rescaling process, prior to sampling the posterior distributions, is crucial to get more realistic statistical uncertainties on the values of the model parameters.

The problem with the image multiplicity and higher $\chi^2$ value indicated that the initial cluster mass model was not flexible enough to reproduce the observed multiple image positions. This is not unexpected, as we included several more families with some of them at significantly higher redshift \citep[e.g.,][]{Meneghetti17}.

With the initial setup, the main contribution to the $\chi^2$ value came from the triple 301 and the quad 401 families (see Fig.~\ref{fig:HFF_zoom}). Because of this, we tested different possibilities for slightly increasing the model flexibility around the northern galaxy group. In the following, we describe a representative selection of them.  For the model selection, we used the Bayesian information criterion \citep[BIC;][]{schwarz78_BIC}, Akaike information criterion \citep[AIC;][]{akaike74_AIC}, and AIC with correction for small sample sizes \citep[AICc;][]{sugiura78_AICc}, under the assumption of the same multiple-image positional uncertainty for all tested models.

First, since the DM halo associated with that specific galaxy group has been neglected so far, we included a fourth PIEMD profile to mimic the DM halo $H_\text{4N}$, either with a fully free position (i.e., a prior range of $[0\arcsec, 60\arcsec]$ for both $x_\text{H}$ and $y_\text{H}$) or forced to be within the group (through a smaller prior range). In both cases, we only allowed the values of the position and the Einstein radius to vary and required the halo to be spherical. This additional DM halo $H_\text{4N}$, with only three free parameters, helped significantly to reproduce the image positions of systems 301 and 401.

We further tested the effect of including, instead of $H_\text{4N}$, two significantly elongated cluster members (marked with filled red circles in Fig.~\ref{fig:HFF_zoom}) of this group outside of the scaling relations. This means that those two cluster members have free Einstein and truncation radii, as well as a free axis ratio and position angle. Despite the relatively high number of additional free parameters (eight in total), we only achieved a minor improvement. Supported by the BIC, AIC and AICc values, we rejected this variation of the lens mass parametrization.

As an alternative, we exploited the multi-plane \citep{blandford86, schneider06} capabilities of \GLEE\ \citep[][see e.g., \citealt{chirivi18, schuldt19}; Acebron et al. (in prep) for detailed application examples]{suyu10a_GLEE, suyu12b_GLEE} and considered the effect of including, instead of $H_\text{4N}$, three galaxies that do not belong to the cluster (marked in Fig.~\ref{fig:HFF_zoom}), but which may significantly affect the lensing potential at their specific redshifts (galaxy 10969 at $z=0.9759$, galaxy 10875 at $z=0.2239$, and galaxy 11585 at $z=0.7495$, see Table~\ref{tab:z_muse}). One dPIE profile, with free Einstein and truncation radii, for each of these three galaxies was chosen to represent their total mass contribution. To limit the number of additional free parameters (six in total), we assumed that the profiles are spherical. This also resulted in an improvement compared to the initial parameterization. However, considering the additional free parameters, this is only a minor improvement and not as significant as with $H_\text{4N}$, which requires only three new parameters. 

Moreover, instead of including $H_\text{4N}$ around the northern galaxy group, we also test the possibility of locating it at the galaxy group position in the north-east, specifically ($-17.0510\arcsec, 101.1450\arcsec$) away from the BCG. We fixed the position here but allow the halo to be elliptical, resulting in the same number of additional free parameters (three in total) as for the model with $H_\text{4N}$ near the multiple image families 301 and 401. However, this configuration is not as good as that with $H_\text{4N}$. We further considered models containing both $H_\text{4N}$ and the halo in the north-east, namely, five DM halos in total. 

Specifically, we also tried a model that includes the northern halo $H_\text{4N}$ but with variable axis ratio, position angle, and core radius. In addition, we included the halo located at the north-east with free axis ratio, position angle, and Einstein and core radii, as a fifth DM halo, as well as the two cluster members outside of the scaling relations. This sums up to 15 additional free parameters compared to our final model, resulting in a number of degrees of freedom of 95 instead of 110. With this parameterization, we obtained a RMS value of 0.37\arcsec instead of \rms. Although this is a notable difference in the achieved RMS value, it does not justify 15 additional parameters based on the BIC, AIC, and AICc values. As a consequence, we finally selected the parametrization with just $H_\text{4N}$ as our final model, which we discuss in more detail below.

\begin{table}[t!]
    \caption{Main characteristics of our final cluster mass model in comparison to the model from \citetalias{grillo16}.}
    \label{tab:overview}
    \begin{center}
    \begin{tabular}{cccc}
    \hline \hline \noalign{\smallskip}
    Component & Property & \multicolumn{2}{c}{Amount}  \\ \hline 
    & & \citetalias{grillo16} & This work \\ \hline
    \multirow{5}{*}{Observables} & Multiple images & 88 & 106 \\
    & families & 28 & 34 \\
    & Redshift range & [1.24, 3.70]& [1.24, 5.98] \\
    & cm (spec-$z$) & 164 & 195 \\
    & cm (photo-$z$) & 136 & 113 \\ \hline
    \multirow{8}{*}{Model} & cm (fixed) & 288 & 306 \\
    & cm (free) & 2 &2 \\ 
    & DM halos ($q_\text{H}\not=1$) & 3 & 3  \\
    & DM halos ($q_\text{H}\equiv1$) & 0 & 1  \\
    & free mass parameters & 28 & 31\\
    & free $z$ parameters & 3 & 3 \\
    & d.o.f. & 89 & 110 \\
    & RMS & 0.26\arcsec & \rms \\ \hline
    \end{tabular}
    \end{center}
    \textbf{Notes.} Cluster members (cm) are divided into those with Einstein and truncation radii following the scaling relations (fixed) and those outside these relations (free). The fixed cluster members are assumed to be spherical. In the text, we refer to the spherical DM halo as $H_\text{4N}$.
\end{table}

\subsection{Results of the final model}
\label{sec:modeling:results}

{\renewcommand{\arraystretch}{1.2}
\begin{table}[t!]
    \caption{Model parameter estimates, as the median values, with 68\% CL (i.e., $1\sigma$) uncertainties and the adopted prior ranges.}
    \label{tab:modelparameter}
    \begin{center}
    \begin{tabular}{cccc}
    \hline \hline \noalign{\smallskip}
    Profile & Parameter & Prior & Median $\pm 1 \sigma$ \\ \hline
    \multirow{6}{*}{1$^\text{st}$ Halo} & $x_\text{H} ~[\arcsec]$   & $[-60,60]$ &  $ -0.5 \pm  0.3 $ \\
            & $y_\text{H} ~[\arcsec]$ & $[-60,60]$ & $ 0.40_{- 0.20}^{+ 0.21} $ \\
            & $q_\text{H}$ & $[0.2,1]$  & $ 0.507_{- 0.020}^{+ 0.015} $  \\
            & $\phi_\text{H}$ [rad]& $(-$$\infty, +\infty ) $ & $ 0.60 \pm  0.02 $ \\
            & $\theta_\text{E,H} ~[\arcsec]$ & $[0, +\infty ) $& $ 31.5_{- 3.0}^{+ 3.1} $ \\
            & $r_\text{c,H} ~[\arcsec]$ & $[0, +\infty ) $ & $ 8.72_{- 0.84}^{+ 0.90} $ \\ \hline
    \multirow{6}{*}{2$^\text{nd}$ Halo} & $x_\text{H} ~[\arcsec]$   & $[-60,60]$ & $ -22.6_{- 2.9}^{+ 1.6} $  \\
            & $y_\text{H} ~[\arcsec]$ & $[-60,60]$ & $ -26.8_{- 2.9}^{+ 1.8} $ \\
            & $q_\text{H}$ & $[0.2,1]$  & $ 0.67_{- 0.20}^{+ 0.11} $ \\
            & $\phi_\text{H}$ [rad]& $(-$$\infty, +\infty ) $ & $ 0.53_{- 0.35}^{+ 0.15} $ \\
            & $\theta_\text{E,H} ~[\arcsec]$ & $[0, +\infty ) $& $ 8.52_{- 1.65}^{+ 1.80} $ \\
            & $r_\text{c,H} ~[\arcsec]$ & $[0, +\infty ) $ & $ 7.1 \pm  1.8 $\\ \hline
    \multirow{6}{*}{3$^\text{rd}$ Halo} & $x_\text{H} ~[\arcsec]$   & $[-60,60]$ & $ 25.5_{- 1.0}^{+ 0.7} $ \\
            & $y_\text{H} ~[\arcsec]$ & $[-60,60]$ & $ 47.60_{- 0.25}^{+ 0.24} $ \\
            & $q_\text{H}$ & $[0.2,1]$  & $ 0.217_{- 0.014}^{+ 0.030} $ \\
            & $\phi_\text{H}$ [rad]& $(-$$\infty, +\infty ) $ & $ 0.60 \pm  0.04 $ \\
            & $\theta_\text{E,H} ~[\arcsec]$ & $[0, +\infty ) $& $ 9.8 \pm  0.9 $ \\
            & $r_\text{c,H} ~[\arcsec]$ & $[0, +\infty ) $ & $ 0.93_{- 0.35}^{+ 0.40} $ \\ \hline      
    \multirow{4}{*}{4$^\text{th}$ Halo ($H_\text{4N}$)} & $x_\text{H} ~[\arcsec]$   & $[0,60]$ & $ 15.23_{- 0.40}^{+ 0.35} $  \\
            & $y_\text{H} ~[\arcsec]$ & $[0,60]$ & $ 59.6_{- 0.6}^{+ 0.3} $ \\
            & $\theta_\text{E,H} ~[\arcsec]$ & $[0, +\infty ) $& $ 4.5 \pm  0.5 $ \\ \hline
    \multirow{2}{*}{Scaling relation} & $\theta_\text{E,BCG}$ [\arcsec] & $[0, + \infty )$ & $ 1.57 \pm  0.24 $ \\
            & $r_\text{t,BCG}$ [\arcsec] & $[0, +\infty)$ & $ 15.8_{- 5.9}^{+ 13.0} $\\ \hline
    \multirow{4}{*}{1$^\text{st}$ member} & $q$ & $[0.2,1]$  & $ 0.8_{- 0.3}^{+ 0.2} $ \\
            & $\phi$ [rad]& $(-$$\infty, +\infty ) $ & $ 1.45_{- 0.82}^{+ 1.05} $ \\
            & $\theta_\text{E} ~[\arcsec]$ & $[0, +\infty ) $& $ 1.85_{- 0.83}^{+ 6.1} $ \\
            & $r_\text{t} ~[\arcsec]$ & $[0, +\infty ) $ & $ 1.04_{- 0.83}^{+ 2.45} $\\ \hline
    \multirow{4}{*}{2$^\text{nd}$ member} & $q$ & $[0.2,1]$  & $ 0.6 \pm  0.3 $ \\
            & $\phi$ [rad]& $(-$$\infty, +\infty ) $ & $ 1.5_{- 1.0}^{+ 1.2} $ \\
            & $\theta_\text{E} ~[\arcsec]$ & $[0, +\infty ) $& $ 2.2_{- 1.1}^{+ 4.6} $ \\
            & $r_\text{t} ~[\arcsec]$ & $[0, +\infty ) $ & $ 0.39_{- 0.30}^{+ 0.71} $\\ \hline
    \multirow{3}{*}{photo-$z$} & $z_6$ & $[0,6]$  & $ 2.76_{- 0.15}^{+ 0.20} $ \\
            & $z_7$ & $[0,6]$  & $ 2.74 \pm  0.06 $ \\
            & $z_8$ & $[0,6]$  & $ 2.84_{- 0.05}^{+ 0.06} $\\ \hline 
    \end{tabular}
    \end{center}
    \textbf{Notes.} The positions are given relative to the BCG, Einstein radii are given for a source redshift of infinity, and all position angle values, measured counterclockwise from the positive $x$-axis, are converted to the range $[0, \pi]$. For the prior ranges, square brackets indicates bounds included in the prior range, while round brackets indicate bounds excluded. The median values and $1\sigma$ uncertainties are computed by excluding the first 20,000 accepted iterations as burn-in points. 
\end{table}
}

Thanks to the additional halo $H_\text{4N}$ and significant sampling sequences, we were able to achieve a final RMS value of \rms, which is notably lower than that with the initial cluster total mass parametrization and the new multiple images. This indicates that the additional halo $H_\text{4N}$, with only three free parameters, is amply justified. We provide in Table~\ref{tab:overview} a summary of the main characteristics of that model, in comparison to the model presented by \citetalias{grillo16}. In Table~\ref{tab:modelparameter}, we list the parameter estimates of the final model, given as the median values with statistical uncertainties defined by the 16$^\text{th}$ and 84$^\text{th}$ percentiles (corresponding to the 68\% confidence intervals). 


\begin{figure}[t!]
    \centering
    \includegraphics[trim={10 15 0 10},clip, width=0.99\linewidth]{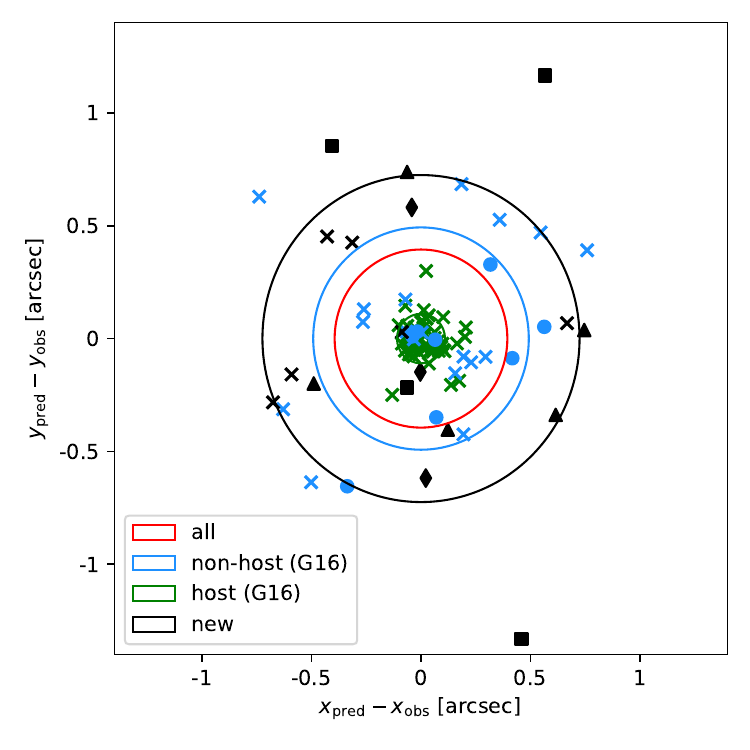}\\
    \includegraphics[trim={0 15 10 10},clip, width=0.99\linewidth]{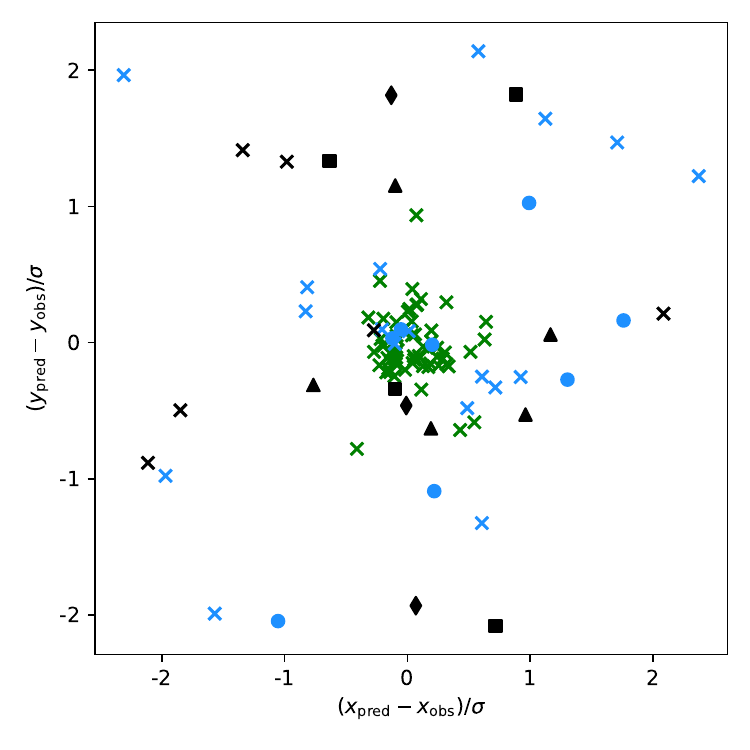}
    \caption{Separation between predicted and observed image positions of the multiple images (top) and positional differences scaled by the corresponding uncertainty (bottom) for our final reference model. The multiple images are grouped into images (1) belonging to the SN host (green), and at different redshifts, both from (2) \citetalias{grillo16} (blue) and (3) this work (black) listed in Table~\ref{tab:sources}. We highlight the multiple images with variable redshift (dots), those which are only identified by \MUSE \, (squares for family 401 and triangles otherwise), and family 301 (diamonds). In the top panel, we further indicate the corresponding RMS values and the total RMS of \rms \, (red) via the radii of the circles.}
    \label{fig:rms}
\end{figure}

\begin{figure*}[t!]
    \centering
    \includegraphics[trim={80 45 35 30},clip, width=0.85\linewidth]{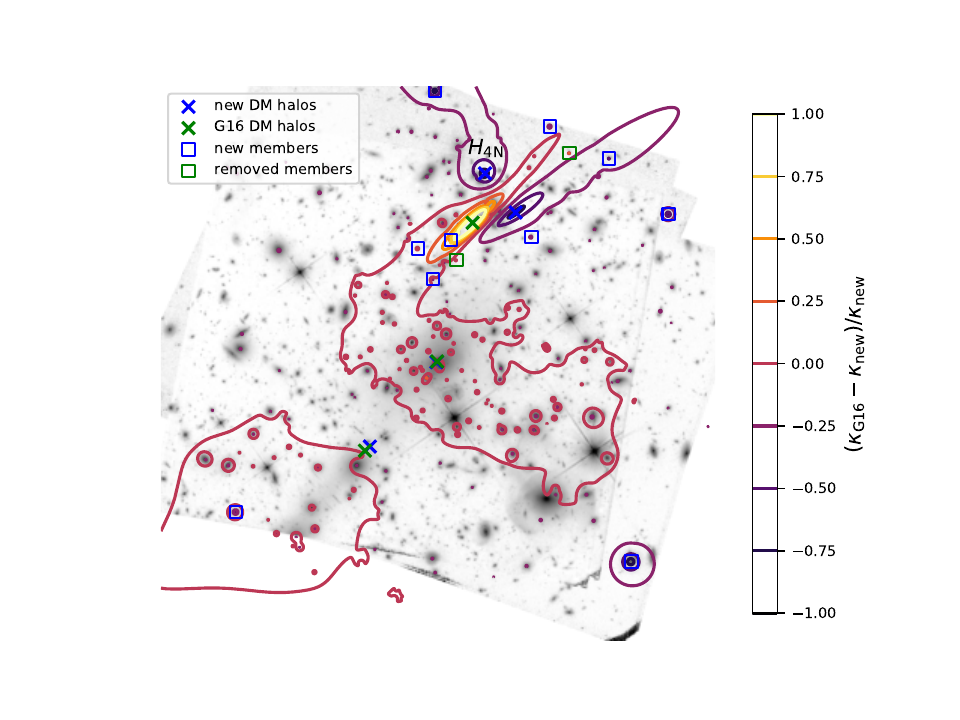}
    \caption{Relative difference between the convergence map from \citetalias{grillo16} and our final total mass model, indicated by color-coded contour levels. We highlight the positions of new (blue) and \citetalias{grillo16}/removed mass density profiles (green), for the DM halos (crosses) and cluster members (squares). The spherical halo $H_\text{4N}$ is also labeled. For an improved orientation, we show in the background the red channel of the HFF color image (see Fig.~\ref{fig:HFF}).}
    \label{fig:kappa}
\end{figure*}

Figure~\ref{fig:rms} shows the contribution to the final RMS of the newly identified multiple images (black) compared to those of the SN host (green) and the non-host images from \citetalias{grillo16} (blue). In detail, the top panel shows the difference between the model-predicted and observed image positions, with circle radii indicating the corresponding RMS values. This illustrates the significant contribution to the total RMS of the newly identified images, driven mainly by the quad 401 from the northern galaxy group. We recall that this system is only detected in the \MUSE \, datacube, such that we adopt for its images a factor 2 larger position uncertainties $(\Delta x, \Delta y)$, compared to that adopted for the images identified in the \HST \, data. Thus, as shown in the bottom panel, the effective contribution of this system to the total $\chi^2$ value, which is the quantity we minimized, is on a similar level compared to other families at different redshifts. We further note that the comparison of the individual $\chi^2$ contribution of each family is difficult, as it depends on the number of their multiple images (e.g., double, triple, quad), on the image distances to the well-constrained regions around the SN Refsdal host images, as well as on their redshift. In addition, three families (IDs 6, 7, and 8; see Table~2 in \citetalias{grillo16}) of the \citetalias{grillo16} non-host group (blue) are only photometrically confirmed. As noted above, we optimized their redshifts in the range of $[0,6]$, making it easier to reproduce those image positions. This leads to a lower RMS value for that group of multiple images. 


Furthermore, Fig.~\ref{fig:rms} highlights how well the images of the SN Refsdal host are reproduced, thanks to a large number of different knots (see Table~2 of \citetalias{grillo16}). This implies a good reconstruction of the lens Fermat potential around the multiple images of SN Refsdal, which is crucial for precise and accurate time-delay cosmography, as shown by \citet{grillo18, grillo20, grillo24}.

Moreover, we show in Fig.~\ref{fig:kappa} a direct comparison with the best-fitting model of \citetalias{grillo16}, in the  form of a $\kappa$-map ratio. In detail, we show the ratio of the difference between the $\kappa_\text{G16}$ map from \citetalias{grillo16} and the map from our final model $\kappa_\text{new}$, and $\kappa_\text{new}$, namely, $\frac{\kappa_\text{G16}(x,y)-\kappa_\text{new}(x,y)}{\kappa_\text{new}(x,y)}$ .


As expected, the two maps are very similar around the cluster core, where we have most of the observables, while we find larger differences in the north, where we added the additional halo $H_\text{4N}$ and several new multiple images. We note that the third halo changes its center to compensate for the new mass component. As a reference, we marked the positions of the different halos with a cross, both at the new locations (blue) and the positions published in \citetalias{grillo16} (green). The positions of the two removed cluster members (green squares) and ten added cluster members (blue squares) are indicated as well, highlighting only minor differences in the $\kappa$ map. Moreover, the switch from the CLASH to the HFF magnitudes used in the scaling relations explains small deviations at the individual cluster member positions. However, these small differences are expected to have a minor influence on the cluster total mass distribution, which is supported by the well-reproduced image positions of the SN Refsdal host (c.~f. Fig.~\ref{fig:rms}).

As a further test, we reconstructed with \GLEE \, the unlensed surface brightness distribution of the SN Refsdal host on a $75\times75$ pixel grid. This reconstruction is then mapped back to the image plane to obtain the lensed image predicted by the model, which can be directly compared to the observed image. It results in similar residuals, as shown in Fig.~7 of \citetalias{grillo16} and Fig.~5 of \citet{grillo20}, which is in agreement with the minor differences in the $\kappa$-map reconstruction.

Moreover, we estimated the model-predicted time delays for the multiple images of SN Refsdal and found an agreement well within $1\sigma$ compared to the previously predicted values from \citetalias{grillo16}. This demonstrates the robustness of the lensing model around SN Refsdal, thanks to the high number of identified knots of its host galaxy. This comparison is intended as a consistency check of this new model, improved with additional multiple images, but not suited for time-delay cosmography applications. Thus, we refrained from releasing the estimates and refer to future publications for new robust time-delay predictions and cosmological results.

\FloatBarrier
\section{Discussion and conclusion}
\label{sec:conclusion}

Together with this work, we have released a catalog containing \musecat \, new, secure spectroscopic redshifts obtained from our recent 5h \MUSE \, observations covering a galaxy group north of the lensing cluster MACS~1149. By combining these new redshifts with deep \HST \, images and a re-analysis of the \MUSE \, data cube presented by \citetalias{grillo16}, we identified 17 new and spectroscopically confirmed multiple images, belonging to six different families. This has resulted in 106 secure multiple images from 34 different sources or knots of the SN Refsdal host. It has allowed us to nearly double the number of sources at different redshifts compared to previous publications, and to extend the redshift range by a factor of about 2.

Following previous models of this cluster, we have considered as cluster members in our total mass model all galaxies within the redshift range of $0.52 \leq z \leq 0.57$, down to a magnitude limit of $m_\text{F160W} \leq 24$. We favored the deep HFF image over the CLASH image used in \citetalias{grillo16}, and released here all measured magnitudes. The new photometric and spectroscopic observations have resulted in the rejection of two previously identified members and the identification of ten new ones, all spectroscopically confirmed by either GLASS \citep{treu16} or \MUSE \, (\citetalias{grillo16} or this work). Furthermore, with our new redshift catalog, we have confirmed 22 cluster members that had previously only been photometrically selected. In total, we now have 308 secure cluster members, of which 195 (63.3\%) have been now spectroscopically confirmed.

The new \MUSE \, observations have allowed us to build an enhanced strong lensing model of MACS~1149. We tested several different total mass  parametrizations and obtained our final model, which best reproduces the observed multiple image positions with a limited number of free parameters and a RMS value of \rms. This level of accuracy is well in agreement with similar strong lensing models of various other clusters \citep[e.g.,][]{bergamini19, bergamini21, caminha23}. In this final cluster mass model, which is publicly available\footnote{Beside providing the mass density parametrization and the corresponding obtained median values in Table~\ref{tab:modelparameter}, we release the model at the following link upon publication: \url{https://www.fe.infn.it/astro/lensing/} \label{footnote:link}}, we have included 308 cluster members each with a dPIE profile. For 306 members, the Einstein and truncation radii are computed through the scaling relations. Moreover, the cluster mass distribution on a larger scale, consisting mainly of the dark matter component of the galaxy cluster and galaxy group, was modeled with four PIEMD profiles, one of which is assumed to be spherical. This has resulted in a total of 34 optimized parameters, along with a $\chi^2$ value of around 110 (for a multiple-image positional uncertainty of $\Delta x = \Delta y = 0.32\arcsec$), corresponding to the degrees of freedom of the presented model.

While exploiting and presenting extensive spectroscopic data from \MUSE \, and imaging data from \HST, this cluster has been recently observed by the \JWST \, (PIs N.~Luetzgendorf,  M.~Stiavelli,  F.~Sun,  C.~Willott) to complement the existing data of this unique lens. These observations will reveal details of several high redshift sources, including the object called JD1 \citep{hashimoto18} located in the north-west of MACS~1149 at a redshift of $z=9.1$, as presented by \citet{hashimoto18} and \citet{stiavelli23}. Although we have extended our total mass model further to the north with the presented \MUSE \, observations, the model still lacks constraints in the region around JD1. Further details of this extreme and peculiar source -- or sources -- (see Fig.~1 of \citealt{stiavelli23}) are required to better understand its physical properties.

The \JWST \, NIRSpec and NIRCam observations will allow us to reveal and spectroscopically confirm additional multiply lensed sources and to identify more knots in known multiple images. With our presently improved total mass model of MACS~1149, we have paved the way to incorporating these \JWST \, data. Moreover, while this model is optimized using the observed multiple image positions only, it will enable us to build an extended strong lensing model that exploits the deep \HST \, (and \JWST) images through the reconstruction of the surface brightness distribution of the SN Refsdal host galaxy. This will increase the number of observables around the multiple images of SN Refsdal by around two orders of magnitude and offer the unique opportunity to demonstrate and visualize the accuracy of the cluster total mass model. This will also be crucial to improving the cosmographic measurements (see e.g., \citealt{grillo24}) with the time delays of supernovae strongly lensed by galaxy clusters.

\begin{acknowledgements}
We thank Ana Acebron and the anonymous referee for helpful discussions and comments. We acknowledge financial support through grants PRIN-MIUR 2017WSCC32 and 2020SKSTHZ. SS has received funding from the European Union’s Horizon 2022 research and innovation programme under the Marie Skłodowska-Curie grant agreement No 101105167 - FASTIDIoUS. AM acknowledges financial support through grant NextGenerationEU" RFF M4C2 1.1 PRIN 2022 project 2022ZSL4BL INSIGHT. SHS thanks the Max Planck Society for support through the Max Planck Fellowship. This project has received funding from the European Research Council (ERC) under the European Union's Horizon 2020 research and innovation programme (LENSNOVA: grant agreement No 771776). This research is supported in part by the Excellence Cluster ORIGINS which is funded by the Deutsche Forschungsgemeinschaft (DFG, German Research Foundation) under Germany's Excellence Strategy -- EXC-2094 -- 390783311.

This research is based on observations made with the NASA/ESA \textit{Hubble} Space Telescope obtained from the Space Telescope Science Institute, which is operated by the Association of Universities for Research in Astronomy, Inc., under NASA contract NAS 5-26555. Based on observations collected at the European Southern Observatory under ESO programmes 294.A-5032 and 105.20P5.001.

Software Citations:
This work uses the following software packages:
\href{https://github.com/astropy/astropy}{\texttt{Astropy}}
\citep{astropy1, astropy2},
\href{https://github.com/dfm/emcee}{\texttt{Emcee}}
\citep{emcee},
\href{https://github.com/matplotlib/matplotlib}{\texttt{matplotlib}}
\citep{matplotlib},
\href{https://github.com/numpy/numpy}{\texttt{NumPy}}
\citep{numpy1, numpy2},
\href{https://www.python.org/}{\texttt{Python}}
\citep{python},
\href{https://github.com/scipy/scipy}{\texttt{Scipy}}
\citep{scipy}.
\end{acknowledgements}

\bibliographystyle{aa}
\bibliography{main}

\appendix

\section{MUSE redshift catalog}

We release in Table~\ref{tab:z_muse} the full MUSE\, redshift catalog (see Sect.~\ref{sec:muse} for details on the data reduction and redshift extraction). We divide into foreground, cluster members, and background objects. This catalog is also online$^{\ref{footnote:link}}$ available together with our strong lensing model quantities. 


\begin{table}[ht!]
    \caption{\MUSE \, catalog from new pointing.}
    \label{tab:z_muse}
\end{table}
\tablefirsthead{
\toprule ID& R.A. & Decl. & $z_{\rm sp}$ & QF \\  & (J2000) & (J2000) & & \\ \midrule \midrule}
\tablehead{%
\multicolumn{5}{c}%
{{\bfseries  Table~\ref{tab:z_muse} Continued:} MUSE catalog.} \\
\toprule
ID& R.A. & Decl. & $z_{\rm sp}$ & QF \\  & (J2000) & (J2000) &  & \\ \midrule \midrule}
\begin{xtabular}{ccccc}
\phantom{x}1934 & 177.386230 & 22.402407 & 0.0000 & 3\\
10773 & 177.385284 & 22.415760 & 0.0000 & 3\\
11827 & 177.395355 & 22.410061 & 0.0000 & 3\\
\phantom{xx}175 & 177.402374 & 22.407394 & 0.0000 & 3\\
10875 & 177.392548 & 22.412317 & 0.2239 & 3\\
11239 & 177.391602 & 22.413786 & 0.2405 & 9\\
12173 & 177.386124 & 22.408899 & 0.2658 & 9\\
11010 & 177.386307 & 22.416983 & 0.3707 & 3\\
13059 & 177.397049 & 22.404455 & 0.4246 & 3\\
11899 & 177.400131 & 22.410467 & 0.4300 & 3\\
12148 & 177.392380 & 22.409252 & 0.4392 & 3\\
12927 & 177.389771 & 22.405228 & 0.4945 & 3\\
13125 & 177.395447 & 22.404039 & 0.5139 & 3\\
13578 & 177.392883 & 22.403997 & 0.5142 & 3\\ \hline
\phantom{xxx}77 & 177.386322 & 22.403538 & 0.5262 & 2\\
\phantom{x}7566 & 177.390961 & 22.401681 & 0.5274 & 3\\
10943 & 177.383530 & 22.414936 & 0.5277 & 3\\
10872 & 177.394714 & 22.411367 & 0.5282 & 3\\
10749 & 177.382568 & 22.416515 & 0.5290 & 3\\
10825 & 177.392563 & 22.410583 & 0.5303 & 3\\
11562 & 177.388306 & 22.411697 & 0.5303 & 3\\
11205 & 177.386620 & 22.413649 & 0.5311 & 3\\
13285 & 177.391037 & 22.403284 & 0.5323 & 2\\
11547 & 177.401596 & 22.411303 & 0.5336 & 3\\
12244 & 177.400696 & 22.408604 & 0.5337 & 2\\
12955 & 177.403061 & 22.404383 & 0.5339 & 3\\
\phantom{xx}250 & 177.390732 & 22.409433 & 0.5340 & 2\\
11888 & 177.393356 & 22.410439 & 0.5341 & 3\\
10466 & 177.391098 & 22.404907 & 0.5343 & 3\\
10542 & 177.391830 & 22.405287 & 0.5343 & 3\\
11038 & 177.394730 & 22.414410 & 0.5347 & 3\\
11202 & 177.386383 & 22.413706 & 0.5352 & 3\\
12411 & 177.391342 & 22.407789 & 0.5360 & 3\\
10412 & 177.398193 & 22.416279 & 0.5362 & 3\\
11009 & 177.388107 & 22.413761 & 0.5372 & 3\\
11922 & 177.396805 & 22.410011 & 0.5379 & 3\\
13521 & 177.393814 & 22.402306 & 0.5381 & 3\\
12783 & 177.398453 & 22.405361 & 0.5382 & 3\\
11974 & 177.383591 & 22.410727 & 0.5382 & 3\\
10838 & 177.397720 & 22.415215 & 0.5388 & 3\\
11951 & 177.394073 & 22.412720 & 0.5391 & 3\\
10492 & 177.383721 & 22.417095 & 0.5395 & 3\\
11618 & 177.397034 & 22.411560 & 0.5406 & 3\\
12791 & 177.399261 & 22.405941 & 0.5418 & 3\\
11559 & 177.388031 & 22.411810 & 0.5418 & 3\\
12377 & 177.394043 & 22.408434 & 0.5421 & 3\\
13256 & 177.392105 & 22.403413 & 0.5423 & 2\\
10787 & 177.390106 & 22.415707 & 0.5423 & 3\\
12629 & 177.384003 & 22.406673 & 0.5429 & 3\\
13006 & 177.397980 & 22.404575 & 0.5433 & 2\\
13114 & 177.390152 & 22.403889 & 0.5443 & 3\\
11346 & 177.392942 & 22.411877 & 0.5445 & 3\\
11468 & 177.399490 & 22.412281 & 0.5471 & 2\\
10891 & 177.382751 & 22.413095 & 0.5474 & 3\\
12443 & 177.394272 & 22.407550 & 0.5474 & 3\\
12242 & 177.398163 & 22.407408 & 0.5475 & 3\\
10247 & 177.403717 & 22.404572 & 0.5476 & 3\\
10743 & 177.383163 & 22.415918 & 0.5478 & 3\\
12261 & 177.401642 & 22.409443 & 0.5480 & 2\\
13133 & 177.397919 & 22.403936 & 0.5486 & 3\\
11722 & 177.387070 & 22.411036 & 0.5495 & 3\\
10946 & 177.395233 & 22.414932 & 0.5501 & 3\\
10766 & 177.398132 & 22.415958 & 0.5506 & 2\\
10688 & 177.398300 & 22.416018 & 0.5518 & 2\\
10865 & 177.384674 & 22.414953 & 0.5518 & 2\\
12169 & 177.388123 & 22.408403 & 0.5529 & 3\\
12900 & 177.389465 & 22.406374 & 0.5531 & 3\\
11759 & 177.383896 & 22.410284 & 0.5537 & 3\\
11162 & 177.390289 & 22.413977 & 0.5540 & 2\\
10236 & 177.395798 & 22.418215 & 0.5542 & 2\\
10846 & 177.399804 & 22.414903 & 0.5546 & 3\\
10217 & 177.396835 & 22.417459 & 0.5550 & 3\\
12379 & 177.384369 & 22.409304 & 0.5550 & 3\\
10239 & 177.396912 & 22.418097 & 0.5553 & 3\\
10233 & 177.392883 & 22.418200 & 0.5561 & 3\\
11341 & 177.393127 & 22.411263 & 0.5567 & 3\\
10760 & 177.397675 & 22.416185 & 0.5571 & 3\\
10144 & 177.399002 & 22.418392 & 0.5576 & 3\\
\phantom{xx}139 & 177.403580 & 22.405603 & 0.5582 & 2\\
\phantom{xx}284 & 177.396820 & 22.410732 & 0.5584 & 2\\
10888 & 177.399429 & 22.415444 & 0.5589 & 3\\
12099 & 177.389938 & 22.409636 & 0.5592 & 3\\
12096 & 177.397568 & 22.409370 & 0.5606 & 2\\
11151 & 177.396591 & 22.413328 & 0.5607 & 3\\
11372 & 177.393830 & 22.411690 & 0.5608 & 3\\
11297 & 177.399353 & 22.413298 & 0.5609 & 3\\
10993 & 177.392181 & 22.414761 & 0.5612 & 3\\
11289 & 177.394211 & 22.412466 & 0.5620 & 3\\
11671 & 177.398270 & 22.410788 & 0.5645 & 3\\
\phantom{xxx}10 & 177.389145 & 22.401512 & 0.5656 & 9\\
11033 & 177.388870 & 22.414148 & 0.5666 & 3\\
12499 & 177.397049 & 22.407593 & 0.5667 & 3\\ \hline
\phantom{xx}245 & 177.401636 & 22.409258 & 0.741 & 3\\
10845 & 177.384064 & 22.413107 & 0.742 & 3\\
11585 & 177.393600 & 22.410720 & 0.750 & 3\\
10564 & 177.393753 & 22.417341 & 0.821 & 3\\
10579 & 177.398300 & 22.419676 & 0.838 & 3\\
10606 & 177.382324 & 22.415525 & 0.849 & 3\\
12318 & 177.402664 & 22.408005 & 0.851 & 3\\
10789 & 177.395264 & 22.417244 & 0.863 & 3\\
11551 & 177.386093 & 22.411703 & 0.863 & 3\\
11023 & 177.382584 & 22.414490 & 0.886 & 3\\
12845 & 177.401650 & 22.405742 & 0.962 & 3\\
10969 & 177.393440 & 22.412600 & 0.976 & 3\\
\phantom{xx}243 & 177.386627 & 22.409378 & 0.979 & 3\\
\phantom{xxx}45 & 177.387207 & 22.402948 & 1.026 & 3\\
10341 & 177.401001 & 22.404701 & 1.086 & 3\\
\phantom{x}9681 & 177.397049 & 22.421038 & 1.089 & 9\\
12906 & 177.398499 & 22.405746 & 1.089 & 3\\
10678 & 177.395538 & 22.416037 & 1.103 & 3\\
11122 & 177.385842 & 22.413897 & 1.274 & 3\\
11107 & 177.385986 & 22.414076 & 1.274 & 3\\
12028 & 177.383514 & 22.410889 & 1.276 & 3\\
10679 & 177.381729 & 22.416651 & 1.377 & 3\\
10093 & 177.391976 & 22.419028 & 1.482 & 9\\
13021 & 177.400391 & 22.406767 & 1.600 & 3\\
12760 & 177.397018 & 22.406183 & 1.678 & 3\\
12761 & 177.397217 & 22.406183 & 1.678 & 3\\
10972 & 177.392929 & 22.415367 & 1.806 & 2\\
10942 & 177.393921 & 22.413961 & 2.347 & 3\\
10584 & 177.394073 & 22.416553 & 2.488 & 3\\
11775 & 177.383316 & 22.409315 & 2.768 & 3\\
13088 & 177.387924 & 22.404253 & 2.771 & 2\\
13019 & 177.387222 & 22.404619 & 3.017 & 3\\
\phantom{x}2153 & 177.392426 & 22.402571 & 3.130 & 2\\
12424 & 177.401169 & 22.407183 & 3.130 & 3\\
22820 & 177.392715 & 22.403093 & 3.131 & 3\\
13293 & 177.392786 & 22.403189 & 3.131 & 3\\
\phantom{x}2191 & 177.393021 & 22.403545 & 3.132 & 3\\
\phantom{xx}562 & 177.383587 & 22.409872 & 3.188 & 3\\
\phantom{xx}560 & 177.384083 & 22.410684 & 3.188 & 3\\
\phantom{xx}561 & 177.383709 & 22.410443 & 3.188 & 3\\
\phantom{xx}563 & 177.384075 & 22.410203 & 3.188 & 3\\
12338 & 177.401703 & 22.408350 & 3.188 & 3\\
10816 & 177.388123 & 22.415678 & 3.215 & 3\\
\phantom{xx}210 & 177.386322 & 22.408720 & 3.219 & 9\\
\phantom{xx}564 & 177.391847 & 22.417813 & 3.388 & 9\\
301.1 & 177.395027 & 22.412690 & 3.447 & 9\\
301.2 & 177.393511 & 22.413066 & 3.447 & 3\\
301.3 & 177.392831 & 22.412863 & 3.447 & 3\\
\phantom{xx}550 & 177.391884 & 22.408651 & 3.483 & 2\\
\phantom{xx}549 & 177.391557 & 22.408464 & 3.483 & 3\\
\phantom{xx}565 & 177.383499 & 22.414271 & 3.531 & 9\\
\phantom{x}3019 & 177.398224 & 22.414471 & 3.532 & 9\\
10098 & 177.396378 & 22.419118 & 3.574 & 3\\
\phantom{xx}566 & 177.389469 & 22.418364 & 3.597 & 9\\
13261 & 177.391678 & 22.403515 & 3.704 & 3\\
13457 & 177.390839 & 22.402636 & 3.705 & 3\\
\phantom{xx}567 & 177.386664 & 22.417851 & 3.824 & 3\\
\phantom{xx}546 & 177.395420 & 22.415347 & 4.161 & 3\\
202.1 & 177.400121 & 22.404141 & 4.384 & 3\\
202.2 & 177.398323 & 22.403761 & 4.384 & 3\\
12420 & 177.384079 & 22.408073 & 4.734 & 3\\
\phantom{xx}568 & 177.397473 & 22.411118 & 4.845 & 3\\
22804 & 177.388626 & 22.402992 & 5.033 & 9\\
204.1 & 177.400379 & 22.403885 & 5.806 & 9\\
204.2 & 177.399437 & 22.403754 & 5.806 & 9\\
401.1 & 177.391676 & 22.411496 & 5.983 & 9\\
401.2 & 177.395838 & 22.412362 & 5.983 & 3\\ 
401.3 & 177.393575 & 22.411423 & 5.983 & 3\\
401.4 & 177.393185 & 22.413001 & 5.983 & 9\\
\phantom{x}2364 & 177.387711 & 22.405834 & 6.155 & 9\\ \hline \hline
\end{xtabular}\\
\textbf{Note.} Columns give, from left to right, the identification number ID, right ascension, declination, spectroscopic redshift $z_\text{sp}$, and the corresponding uncertainty in terms of a quality flag QF, with 3 as very secure ($\delta z <$ 0.001), 2 as secure ($\delta z <$ 0.01), and 9 as secure (based on a single emission line).\\

\end{document}